\documentclass[10pt,twocolumn,final]{IEEEtran}
\usepackage{amsmath,amssymb,epsfig,color}
\usepackage[breaklinks=true,colorlinks=true,linkcolor=black,urlcolor=dblue,citecolor=black,pdfpagemode=None,pdfstartview=FitH]{hyperref}
\usepackage{graphicx,verbatim}

\newtheorem{definition}{Definition}
\newtheorem{theorem}{Theorem}
\newtheorem{lemma}{Lemma}
\newtheorem{corollary}{Corollary}
\newtheorem{proposition}{Proposition}
\newtheorem{remark}{Comment}[section]
\definecolor{gray}{cmyk}{.2,0.2,.3,.1}
\definecolor{dread}{cmyk}{0,0.9,0.4,0.3}
\definecolor{dblue}{rgb}{0,0,0.5}
\definecolor{dgreen}{rgb}{0,0.3,0}
\definecolor{dgray}{rgb}{0.3,0.3,0}

\newcommand{\xvec}{{\bf x}}
\newcommand{\yvec}{{\bf y}}
\newcommand{\uvec}{{\bf u}}
\newcommand{\vvec}{{\bf v}}
\newcommand{\Rgood}{R_{x1}}
\newcommand{\Rbad}{R_{x2}}
\newcommand{\typ}{A_{\epsilon}^{(n)}}
\newcommand{\styp}{A_{\epsilon}^{*(n)}}
\newcommand{\stypd}{A_{\delta}^{*(n)}}

\newcommand{\tS}{\widetilde{S}}
\newcommand{\ts}{\widetilde{s}}

\newcommand{\mU}{\mathcal{U}}
\newcommand{\mhU}{\mathcal{\hat{U}}}
\newcommand{\mX}{\mathcal{X}}
\newcommand{\mY}{\mathcal{Y}}
\newcommand{\mV}{\mathcal{V}}
\newcommand{\mhV}{\mathcal{\hat{V}}}
\newcommand{\mD}{\mathcal{D}}
\newcommand{\mW}{\mathcal{W}}
\newcommand{\mL}{\mathcal{L}}
\newcommand{\mZ}{\mathcal{Z}}
\newcommand{\Iu}{I\left(U;Y_1,\hat{U}\right)}
\newcommand{\Iv}{I\left(V;Y_2,\hat{V}\right)}

\newcommand{\hW}{\hat{W}}
\newcommand{\hU}{\hat{U}}
\newcommand{\hV}{\hat{V}}
\newcommand{\hu}{\hat{u}}
\newcommand{\hv}{\hat{v}}
\newcommand{\hz}{\hat{z}}
\newcommand{\huvec}{\hat{{\bf u}}}
\newcommand{\hvvec}{\hat{{\bf v}}}

\newcommand{\Perr}{P_e^{(n)}}

\newcommand{\eps}{\epsilon}
\newcommand{\hw}{\hat{w}}

\newcommand{\tend}{\hfill$\blacksquare$}

\title{Broadcast Channels with Cooperating Decoders
\thanks{The authors are with the School of Electrical and Computer
Engineering, Cornell University, Ithaca, NY. URL:
\href{http://cn.ece.cornell.edu}{{\tt http://cn.ece.cornell.edu/}}. Parts of this
work were presented at the International Symposium on Information Theory, Chicago, 2004 and
the International Symposium on Information Theory, Adelaide, Australia, 2005.
Work supported by the National Science Foundation, under awards
CCR-0238271 (CAREER), CCR-0330059, and ANR-0325556.}}
\author{Ron Dabora \hspace{2cm} Sergio D.\ Servetto}

\begin{document}
\maketitle

\begin{picture}(0,0)
\put(0,42){\tt\small To appear in the IEEE Transactions on Information
  Theory, December 2006.}
\end{picture}

\begin{abstract}
We consider the problem of communicating over the general discrete memoryless broadcast channel
(BC) with partially cooperating receivers.  In our setup, receivers
are able to exchange messages over noiseless conference links of
finite capacities, prior to decoding the messages sent from the
transmitter.
 In this paper we formulate the general problem of
broadcast with cooperation. We first find the capacity region for the
case where the BC is physically degraded. Then, we  give
achievability results for the general broadcast
channel, for both the two independent messages case and the single common message case.

\end{abstract}


%%%%%%%%%%%%%%%%%%%%%%%%%%%%%%%%%%%%%%%%%%%%%%%%%%%%%%%%%%%%%%%%%%%%%%%%%%%%%%%%%%%%%%%%%%%%%%%%%%%%%%
%%%%%%%%%%%%%%%%%%%%%%%%%%%%%%%%%%%%%%%%%%%%%%%%%%%%%%%%%%%%%%%%%%%%%%%%%%%%%%%%%%%%%%%%%%%%%%%%%%%%%%

\section{introduction}
\label{sec:intro}

\subsection{Motivation}
In the classic broadcast scenario the receivers decode their
messages independently of each other. However, the increasing interest in
networking motivates the consideration of  broadcast scenarios in
which each node in the network, besides decoding its own
information, tries to help other nodes in decoding. This problem
comes up naturally in sensor networks, where a transmitter external
to the sensor network wants to download data into the network, e.g.,
to configure the sensor array. The concept of cooperation among
receivers is also relevant to general ad-hoc networks, since such
cooperation provides a method for increasing the rates without
increasing the spectrum allocation. Therefore, this motivates the
study of the effect of receiver cooperation on the rates for the
broadcast channel.

\subsection{The Discrete Memoryless Broadcast Channel (DMBC)}
\label{sec:generalBCintro}

The broadcast channel was introduced by Cover in~\cite{Cover:72}.
Following this initial work, Bergmans proved an achievability result
for the degraded BC,~\cite{Bergmans:73}, and also a partial converse
that holds only for the Gaussian broadcast
channel~\cite{Bergmans:74}; in~\cite{Gallager:74} Gallager
established a converse that holds for any discrete memoryless
degraded broadcast channel. In \cite{ElGamal:79} El-Gamal
generalized the capacity result for the degraded broadcast channel
to the ``more capable" case, and in \cite{ElGamal:78} and
\cite{ElGamal:81} he showed that feedback does not increase the
capacity region for the physically degraded case. Several other
classes of broadcast channels were studied in the following years.
For example, the sum and product of two degraded broadcast channels
were considered in \cite{ElGamal:80}, and in \cite{Han:81},
\cite{Pinsker:78} and \cite{Gelfand:77} the deterministic broadcast
channel was analyzed.

For the general broadcast channel, Cover derived an achievable rate region for the case of two independent
senders in \cite{Cover:75}.
In \cite{KornerMarton:77} Korner and Marton considered the capacity of
general broadcast channels with degraded message
sets. The best achievable region and the best upper bound for the two independent senders case
were derived by Marton in~\cite{Marton:79}, and a
simple proof of Marton's achievable region appeared later in~\cite{ElGamalM:81}.
Another upper bound for the general broadcast channel, the so-called degraded, same-marginals (DSM) bound,
was presented in \cite{DSM:2002}. This bound is weaker than the upper bound in \cite{Marton:79} but stronger
than Sato's upper bound previously presented in \cite{Sato:78}.
We note, however, that while Marton's upper bound is the strongest, it is valid only for the two-receiver case,
while Sato's bound and the DSM bound can be extended to more than two receivers.
The effect of feedback on the capacity of the Gaussian broadcast channel was studied in \cite{Ozarow:84} and
\cite{Elia:2004}, and in \cite{HanCosta:87} the case of correlated sources was considered.
A survey on the topic, with extensive references to
previous work, can be found in \cite{Cover:98}.
In recent years the Multiple-Input-Multiple-Output (MIMO) Gaussian broadcast channel
has attracted a lot of attention. Initially, the sum-rate capacity was characterized in \cite{CaireShamai:03},
\cite{VishTse:03}, \cite{JindalGoldsmith:04}, \cite{YuCioffi:04}, and finally, in \cite{Yossi:04} the capacity region
was obtained.

None of the early work on the DMBC considered {\em direct} cooperation between the
receivers.
\begin{figure}[ht]
      \center\scalebox{0.48}{\includegraphics{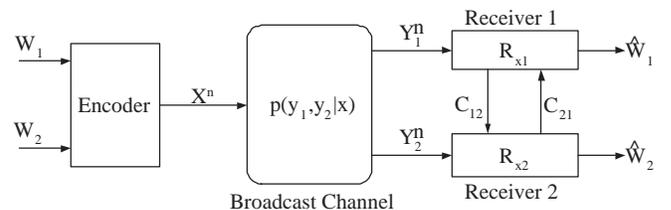}}
  \caption{\small Broadcast channel with two private messages and cooperating receivers.}
  \label{fig:broadcast-cooperation-twousers}
\end{figure}
In the cooperative broadcast scenario, a single transmitter sends
two messages to two receivers encoded in a single channel codeword
$X^n$, where the superscript $n$ denotes the length of a vector.
Each of the receivers gets a noisy version of the codeword,
$Y_1^n$  at $\Rgood$ and $Y_2^n$ at $\Rbad$. After reception, the
receivers exchange messages over noiseless conference links of
finite capacities $C_{12}$ and $C_{21}$, as depicted in Figure
\ref{fig:broadcast-cooperation-twousers}.
%\footnote{The links are
%termed ``finite capacity and noiseless" since a finite capacity link
%is not necessarily noiseless, thus channel coding is required to achieve
%capacity. In the model considered here, treating the links as noiseless implies that
% messages can be exchanged over the links
%with zero probability of error without the use of coding, as long as the rate is below capacity. Such
%model is used for example in \cite{DraperFK:03}, \cite{Zhang:88} and \cite{Willems:83}.}.
 The conference messages are, in general, functions of $Y_1^n$ (at $\Rgood$),
$Y_2^n$ (at $\Rbad$), and the previous messages received from the
other decoder. After conferencing, each receiver decodes its own message.

We note that in a recent work,~\cite{DraperFK:03}, the authors consider
the problem of interactive decoding of a single broadcast message over
the independent broadcast channel by a group of cooperating users. In our work
we extend this scenario to the general channel and also consider the two independent senders~case.

\subsection{Cooperative Broadcast: A Combination of Broadcasting and Relaying}
 The scenario
in which one transceiver helps a second transceiver in decoding a
message is clearly a {\em relay} scenario. Hence, cooperative
broadcast can be viewed as a generalization of the broadcast and
relay scenarios into a hybrid broadcast/relay system, which better
describes future communication networks.

Scenarios of this type have attracted considerable attention recently both from the practical
 and the theoretical aspects. From the practical aspect,
 new protocols are proposed for the collaborative broadcast scenario. For example in
 \cite{Willke:2004} the authors present a protocol for collaborative
 decision making involving broadcasting and relaying.
From the theoretical aspect, there is a considerable effort invested in characterizing
the capacity of an entire network. This work started with \cite{Meulen:71} and recent
results appear in  \cite{GuptaKumar:2000} and the following
work \cite{GastparV:2005}, \cite{GuptaKumar:2003} and \cite{XieKumar:2004}. This work focuses on the Gaussian
case. A complementing approach for studying the performance of a network is to combine the basic
building blocks of a network, namely multiple access, relaying and broadcasting and study the
capacity of these combinations. The recent work on relaying focuses on extending the single relay
results derived in \cite{CoverG:79} to the MIMO case (see for example \cite{WangZhang:2005}) and to the multiple
level case \cite{XieKumar:2005}, \cite{KramerGupta:2003}. Another recent result
was introduced in \cite{MOT:05} where joint decoding was applied to the combined decode-and-forward and
estimate-and-forward scheme of \cite[theorem 7]{CoverG:79}. A third approach for studying the
performance of an entire network is the network coding approach sparked by the work of \cite{NetCod:2000},
which focuses on encoding at the nodes for maximizing the network throughput, separately from the
channel coding.

In this paper we focus on the combination of broadcast and relay.
 A relevant work in this context is \cite{Zhang:88},
in which the capacity of a class of independent relay channels with noiseless
relay is derived. Note that the case of
noiseless relay is also related to the Wyner-Ziv problem
\cite{WynerZiv:76}. This relationship  will be highlighted in the
sequel. Lastly, we note that a recent work, \cite{Liang:04}, presented an achievability
result for the general DMBC with a single wireless cooperation channel from
one receiver to the second receiver. This achievable rate region is shown to be the capacity
region for the physically degraded broadcast/relay channel.

\subsection{Main Contributions and Organization}

In the following we summarize the main contributions of this work.
\begin{itemize}
\item We initially study a special case of the general setup
    formulated in Section~\ref{sec:generalBCintro}: the case of the
    physically degraded broadcast channel. Although the physically
    degraded BC is of little practical interest, it is useful in
    developing the coding concept for the general BC with cooperation.
    For the physically degraded BC, we present both an achievability
    result and a converse.  Together, these two results give the capacity region for this setup.
    Furthermore, this new region is shown to be a strict enlargement
    of the classical region without cooperation~\cite{Cover:98}.
\item Next, we give an achievability result for the general BC with
    cooperating receivers. This region is also greater, in general,
    than the classic achievable region given in~\cite{Marton:79} for
    the broadcast channel.
%    We introduce here {\em joint} decoding of source and relay messages at the destination
%    rather than decoding them sequentially as in \cite[theorem 6]{CoverG:79}. We show that the region obtained in
%    this manner is larger than the one obtained by application of \cite[theorem 6]{CoverG:79} to our
%    channel model. We note here that although \cite{MOT:05} used joint decoding in the relay setup, their application did not modify the EAF rate itself, so when specialized to EAF, \cite{MOT:05} recovers \cite[theorem 6]{CoverG:79}. In our work, we apply joint decoding to the EAF itself (for the noiseless relay case), therefore we improve the EAF rate compared to \cite[theorem 6]{CoverG:79}.
\item We also consider
    the case where a single common message is transmitted to both
    receivers. We consider two different cooperation strategies and
    derive the achievable rates for each of them. We also derive an
    upper bound on the achievable rates for this scenario.
    Here we  provide results that explicitly link the available
    cooperation capacity to the increase in the rate of information.
    Lastly, we show that for a special case of the general BC, namely
    when one channel is distinctly better than the other, the upper
    and lower bounds coincide, resulting in the capacity for that
    case.
\end{itemize}

The rest of this paper is structured as follows: in
section~\ref{sec:defs} we define the mathematical framework.  In
section~\ref{sec:degraded} we analyze the physically degraded BC,
and derive the capacity region for that case, and in
section~\ref{sec:generalBC} we present an achievability result for the general broadcast channel
with cooperating receivers.
%Section \ref{sec:general_upper_bound} follows with an upper bound the general channel.
Next, section \ref{sec:commonmsg} presents
%and \ref{sec:upper_bound_common} present an
achievability results and an upper bound on the rates for the case where only a single common message is
transmitted. Concluding remarks are provided in section~\ref{sec:conclude}.

%%%%%%%%%%%%%%%%%%%%%%%%%%%%%%%%%%%%%%%%%%%%%%%%%%%%%%%%%%%%%%%%%%%%%%%%%%%%%%%%%%%%%%%%%%
%%%%%%%%%%%%%%%%%%%%%%%%%%%%%%%%%%%%%%%%%%%%%%%%%%%%%%%%%%%%%%%%%%%%%%%%%%%%%%%%%%%%%%%%%%

\section{Definitions and Notations}
\label{sec:defs}

First, a word about notation: in the following we use $H(\cdot)$ to
denote the entropy of a discrete random variable (RV), and
$I(\cdot;\cdot)$ to denote the mutual information between two
discrete random variables, as defined in \cite[Ch. 2]{cover-thomas:it-book}.
We denote random variables with capital
letters -- $X$, $Y$, etc., and vectors with boldface letters, e.g.,
$\xvec$, $\yvec$. We denote by $\typ(X)$ the weakly typical set for
the (possibly vector) random variable $X$, see \cite[Ch. 3]{cover-thomas:it-book}
for the definition of $\typ(X)$. When
referring to a typical set we may omit the random variables from the
notation, when these variables are clear from the context. We denote
the cardinality of the finite set $\mathcal{A}$ with
$||\mathcal{A}||$. We use $\mX$ to denote the (discrete and finite) range of $X$. Finally, we
denote the probability distribution of the RV $X$ over $\mX$ with $p(x)$
and the conditional distribution of $X$ given $Y$ with $p(x|y)$.

\begin{definition}
A {\em discrete broadcast channel} is a channel with discrete input alphabet
$\mathcal{X}$, two discrete output alphabets, $\mathcal{Y}_1$ and
$\mathcal{Y}_2$, and a probability transition function, $p(y_1,y_2|x)$. We denote
this channel by the triplet $\left(\mathcal{X}, p(y_1,y_2|x), \mathcal{Y}_1 \times \mathcal{Y}_2\right)$.
\end{definition}

\begin{definition}
A {\em memoryless} broadcast channel is a broadcast channel for which
the probability transition function of a sequence of $n$ symbols
is given by $p(y_1^n,y_2^n|x^n) = \prod_{i=1}^np(y_{1,i},y_{2,i}|x_i)$,
where $y_k^n = (y_{k,1}, y_{k,2},...,y_{k,n})$, $k \in \left\{1,2\right\}$, and
$x^n = (x_{1}, x_{2},...,x_{n})$.
\end{definition}

We shall assume the channel to be {\em discrete} and {\em memoryless}.

\begin{definition}
The {\em physically  degraded broadcast channel} is a broadcast channel in which the probability transition function
can be decomposed as $p(y_1,y_2|x) = p(y_1|x)p(y_2|y_1)$. Hence, for the physically
degraded BC we have that $X - Y_1 - Y_2$ form a Markov chain.
\end{definition}

\begin{definition}
{\em An $(R_{12},R_{21})$-conference} between $R_{x1}$ and $R_{x2}$ is defined by two conference message sets
$\mathcal{W}_{12} = \left\{1,2,...,2^{nR_{12}}\right\}$,
$\mathcal{W}_{21} = \left\{1,2,...,2^{nR_{21}}\right\}$, and two mapping
functions, $h_{12}$ and $h_{21}$ which map the received sequence of $n$ symbols
and the conference messages
at one receiver into a message transmitted to the other receiver:
\begin{eqnarray*}
    & h_{12}:& \mathcal{Y}_1^n \times \mW_{21} \mapsto \mathcal{W}_{12}, \\
    & h_{21}:& \mathcal{Y}_2^n \times \mW_{12} \mapsto \mathcal{W}_{21}.
\end{eqnarray*}
We note that this is not the most general definition of a conference, see for example
\cite{Kaspi:85}, \cite {Willems:83} for a more general form. In this paper we consider only
conferences in which each receiver sends at most one message to the other receiver. Note that
there are cases where a single conference message is enough to achieve capacity: for example, in section
\ref{sec:degraded} a single conference step achieves capacity for the physically degraded broadcast channel, and in
\cite {Willems:83} a single conference step achieves capacity for the discrete memoryless multiple access channel
counterpart of the setup discussed here.
\end{definition}

\begin{definition}
A {\em $\left(C_{12},C_{21}\right)$-admissible conference} is a conference for which
$R_{12} \le C_{12}$ and $R_{21} \le C_{21}$.
\end{definition}

\begin{definition}
\label{def:codes}
A $\left(\left(2^{nR_1},2^{nR_2} \right),n,\left(C_{12},C_{21}\right)\right)$ {\em code} for
the broadcast channel with cooperating receivers having conference links of capacities
$C_{12}$ and $C_{21}$ between them, consists of two sets of integers
$\mathcal{W}_1 = \left\{1,2,...,2^{nR_1}\right\}$,
$\mathcal{W}_2 = \left\{1,2,...,2^{nR_2}\right\}$, called {\em message
sets}, an encoding function
\[
    f:\mathcal{W}_1 \times \mathcal{W}_2  \mapsto \mathcal{X}^n,
\]
a $\left(C_{12},C_{21}\right)$-admissible conference
\begin{eqnarray*}
    & h_{12}:& \mathcal{Y}_1^n \times \mW_{21} \mapsto \mathcal{W}_{12}, \\
    & h_{21}:& \mathcal{Y}_2^n \times \mW_{12} \mapsto \mathcal{W}_{21},
\end{eqnarray*}
and two decoding functions
\begin{eqnarray}
\label{eqn:g1_coop}
     & g_1: &  \mathcal{W}_{21} \times \mathcal{Y}_1^n \mapsto \mathcal{W}_1, \\
\label{eqn:g2_coop}
     & g_2: &  \mathcal{W}_{12} \times \mathcal{Y}_2^n \mapsto \mathcal{W}_2.
\end{eqnarray}
\end{definition}

\begin{definition}
\label{def:perr}
The {\em average probability of error} is defined as the probability that
the decoded message pair is different from the transmitted message pair:
\[
   P_e^{(n)} = \mbox{Pr}\left(g_1(W_{21},Y_1^n) \ne W_1\;\; \mbox{or}
               \;\; g_2(W_{12},Y_2^n) \ne W_2  \right).
\]
We also define the average probability of error for each receiver as:
\begin{eqnarray}
\label{eqn:def_pe1_II}
    P_{e1}^{(n)} & = & \mbox{Pr}\left(g_1\left(W_{21},Y_1^n\right) \ne W_1\right), \\
\label{eqn:def_pe2_II}
    P_{e2}^{(n)} & = & \mbox{Pr}\left(g_2\left(W_{12},Y_2^n\right) \ne W_2\right),
\end{eqnarray}
where %$Y_i^k$ denotes the vector $\left(Y_{i,1},Y_{i,2},...,Y_{i,k}\right)$,
%$i\in \{1,2\}$, and
we assume transmission of $n$ symbols for each codeword.
By the union bound we have that $ \max\left\{P_{e1}^{(n)},P_{e2}^{(n)}\right\} \le \Perr \le P_{e1}^{(n)} + P_{e2}^{(n)}$.
 Hence,
$\Perr \rightarrow 0$ implies that both $P_{e1}^{(n)} \rightarrow 0$ and
$P_{e2}^{(n)} \rightarrow 0$, and when both individual error probabilities go to zero then $\Perr$ goes
to zero as well.

In the analysis that follows, we assume that user 1 and user 2 select their respective
messages $W_1$ and $W_2$ independently and uniformly over their respective message sets.
\end{definition}

\begin{definition}
A rate pair $\left(R_1,R_2\right)$ is said to be {\em achievable},
if there exists a sequence of
$\left( \left(2^{nR_1},2^{nR_2} \right),n, \left(C_{12},C_{21}\right) \right)$ codes with
$P_e^{(n)} \rightarrow 0$ as $n \rightarrow \infty$. Obviously, this is satisfied if both
$P_{e1}^{(n)} \rightarrow 0$ and $P_{e2}^{(n)} \rightarrow 0$ as $n$ increases.
\end{definition}

\begin{definition}
The {\em capacity region} for the discrete memoryless broadcast
channel with cooperating receivers is the convex hull of all achievable rates.
\end{definition}

\section{Capacity Region for the Physically Degraded Broadcast Channel with Cooperating Receivers}
\label{sec:degraded}

We consider the physically degraded broadcast channel with three independent messages: a private message
to each receiver and a common message to both.
We note that for the physically degraded channel, following the argument
in \cite[theorem 14.6.4]{cover-thomas:it-book}, we can incorporate a common rate to both receivers
by replacing $R_2$, the private rate to the bad receiver, obtained for the two private messages case with $R_0 + R_2$,
where $R_0$ denotes the rate of the common information.
Without cooperation, the capacity region for the physically degraded BC $X - Y_1 - Y_2$ given in
\cite[theorem 14.6.4]{cover-thomas:it-book}, is the convex hull of all the rate
triplets $(R_0,R_1,R_2)$ that satisfy
\begin{eqnarray}
\label{eqn:Degraded_NoCoope_1}
    R_1 & \le &   I(X;Y_1|U),\\
\label{eqn:Degraded_NoCoope_2}
    R_0 + R_2 & \le &   I(U;Y_2),
\end{eqnarray}
for some joint distribution $p(u)p(x|u)p(y_1|x)p(y_2|y_1)$, where
\begin{equation}
\label{eqn:U_bound_no_coop_degradd}
    ||\mU|| \le \min\left\{ ||\mX||, ||\mY_1||, ||\mY_2||\right\}.
\end{equation}

Next, consider cooperation between receivers over the physically degraded BC.
First note that for this
case, the link from $\Rbad$ to $\Rgood$ does not
contribute to increasing the rates due to
cooperation, and that only the link from $\Rgood$ to $\Rbad$ does.
This is due to the data processing inequality (see \cite[theorem 2.8.1]{cover-thomas:it-book}): since $X - Y_1 -
Y_2$ form a Markov chain, any information about $X$ contained in
$Y_2$ will also be contained in $Y_1$, and thus conferencing
cannot help:
\[
    I(X;Y_1,Y_2) = I(X;Y_1) + \underbrace{I(X;Y_2|Y_1)}_{=\;\;0} = I(X;Y_1).
\]
For the rest of this section then,
we shall consider only a communication link from the good receiver $R_{x1}$,
to the bad receiver $R_{x2}$ (i.e. we set $C_{21} = 0$).
This implies that $W_{21}$ is a constant and we can thus omit it from the analysis.
We begin with a statement of the theorem:

\begin{theorem}
\label{thm:converse}
\it
The capacity region
for sending independent information over the discrete memoryless physically
degraded broadcast channel
$X - Y_1 - Y_2$, with cooperating receivers having a noiseless conference link of capacity $C_{12}$,
as defined in Section~\ref{sec:defs}, is the convex hull of all rate triplets $(R_0,R_1,R_2)$ that satisfy
\begin{eqnarray}
    \label{eqn:phy_deg_coop_R1}
    R_1       & \le & I(X;Y_{1}|U), \\
    \label{eqn:phy_deg_coop_R2}
    R_0 + R_2       & \le & \min\big(I(U;Y_1),I(U;Y_{2}) + C_{12}\big),
\end{eqnarray}
for some joint distribution $p(u)p(x|u)p(y_1,y_2|x)$, where the
auxiliary random variable $U$ has cardinality bounded by
$||\mathcal{U}||\le \min\left\{||\mathcal{X}||,||\mathcal{Y}_1||\right\}$.
\end{theorem}

We note that this result presented in \cite{RonSer:2004} was simultaneously derived in \cite{Liang:04} for
the case of a wireless relay.
%We note that this result, originally presented in \cite{RonSer:2004} was independently derived in \cite{Liang:04} for
%the case of a wireless relay.
%We discuss the proof of theorem \ref{thm:converse} here for two reasons: the
%first is that the benefits of cooperation for this setup are easily seen in the rate constraint for the bad receiver,
%and the second is that the techniques presented in this proof will be used also in the rest of this paper.

\subsection{Achievability Proof}
\label{sec:achieve}

In this section, we show that the rate triplets of theorem~\ref{thm:converse}
are indeed achievable. We will show that the region defined by (\ref{eqn:phy_deg_coop_R1}) and
(\ref{eqn:phy_deg_coop_R2}) with $R_0 = 0$ is achievable. Incorporating $R_0 > 0$ easily follows as explained
earlier.

\subsubsection{Overview of Coding Strategy}

The coding strategy is a combination of a broadcast code as an ``outer" code
 used to split the rate between $R_{x1}$ and $R_{x2}$, and
an ``inner" code for $R_{x2}$, using the code construction for the physically
degraded relay channel, described in~\cite[theorem 1]{CoverG:79}.
We first generate codewords $U^n$ for $R_{x2}$, according to the
relay channel code construction. Then, the codewords for $R_{x2}$ are used as ``cloud centers" for the
codewords transmitted to $R_{x1}$ (which are also the output to the channel).
Upon reception, $R_{x1}$ decodes both its own message and the
message for $R_{x2}$, and then uses the relay code selection to select the message relayed to
$R_{x2}$. $R_{x2}$ uses its received signal, $Y_2^n$, to generate a list of possible $U^n$ candidates, and then
uses the information from $\Rgood$ to resolve for the correct codeword.

\subsubsection{Details of Coding Strategy}

\paragraph{Code Generation}
\begin{enumerate}
\item Consider first the set of $M_R = 2^{nC_{12}}$ relay messages.
These are the messages that the relay $R_{x1}$ transmits to $R_{x2}$ through the
noiseless finite capacity conference link between the two receivers. Index these messages by $s$, where
$s \in \left\{1,2,...,M_R\right\}$.

Next, fix $p(u)$ and $p(x|u)$.

\item For each index $s \in [1,M_R]$, generate $2^{nR_2}$ conditionally independent codewords
${\bf u}(w_2|s) \sim \prod_{i=1}^n p(u_i)$, where $w_2 \in \left\{1,2,...,2^{nR_2}\right\}$.

\item For each codeword $\uvec(w_2|s)$ generate $2^{nR_1}$ conditionally independent
codewords ${\bf x}(w_1,w_2|s) \triangleq
{\bf x}(w_1|\uvec(w_2|s)) \sim \prod_{i=1}^n p(x_i|u_i(w_2|s))$, where $w_1 \in \left\{1,2,...,2^{nR_1}\right\}$.

\item Randomly partition the message set for $R_{x2}$, $\left\{1,2,...,2^{nR_2} \right\}$,
into $M_R$ sets $\left\{S_1, S_2,...,S_{M_R}\right\}$,
by independently and uniformly assigning to each message an index in $\left[1,M_R\right]$.

\end{enumerate}

\paragraph{Encoding Procedure}

Consider transmission of $B$ blocks, each block transmitted using $n$ channel symbols. Here
we use $nB$ symbol transmissions to transmit $B-1$ message pairs
$\left(w_{1,i},w_{2,i} \right) \in \left[1,2^{nR_1}\right]\times \left[1,2^{nR_2}\right]$, $i=1,2,\dots,B-1$.
As $B\rightarrow \infty$ we have that the rate $(R_1,R_2)\frac{B-1}{B} \rightarrow (R_1,R_2)$. Hence, any rate pair achievable
without blocking can be approached arbitrarily close with blocking as well.
Let $w_{1,i}$ and $w_{2,i}$ be the messages intended for $R_{x1}$ and $R_{x2}$ respectively, at the $i$'th block, and
 also assume that $w_{2,i-1} \in S_{s_i}$. $R_{x1}$ has an estimate $\hat{\hat{w}}_{2,i-1}$ of the message sent to
$R_{x2}$ at block $i-1$. Let $\hat{\hat{w}}_{2,i-1} \in S_{\hat{\hat{s}}_i}$.
At the $i$'th block the transmitter outputs the codeword ${\bf x}(w_{1,i},w_{2,i}|s_i)$, and
 $R_{x1}$ sends the index $\hat{\hat{s}}_i$ to $R_{x2}$ through the noiseless conference link.

\paragraph{Decoding Procedure}
Assume first that up to the end of the $(i-1)$'th block there was no decoding error. Hence,
at the end of the $(i-1)$'th block, $R_{x1}$ knows $\left(w_{1,1},w_{1,2},...,w_{1,i-1}\right)$,
$\left(w_{2,1},w_{2,2},...,w_{2,i-1}\right)$ and $\left(s_{1},s_{2},...,s_{i}\right)$, and $R_{x2}$
knows $\left(w_{2,1},w_{2,2},...,w_{2,i-2}\right)$ and $\left(s_{1},s_{2},...,s_{i-1}\right)$. The
decoding at block $i$ proceeds as follows:
\begin{enumerate}
\item $R_{x1}$ knows $s_i$ from $w_{2,i-1}$. Hence, $R_{x1}$ determines uniquely $(\hw_{1,i},\: \hw_{2,i})$ s.t. \\
  $\big(\uvec(\hw_{2,i}|s_i), {\bf x}(\hw_{1,i},\hw_{2,i}|s_i), {\bf y}_1(i)\big) \in  A_{\epsilon}^{(n)}$.  If there is
  none or there is more than one, an error is declared.
\item $R_{x2}$ receives $s_i$ from $R_{x1}$. From knowledge of $s_{i-1}$ and ${\bf y}_2(i-1)$, $R_{x2}$ forms a
list of possible messages, $\mathcal{L}(i-1) = \left\{w_{2}: \left({\bf y}_2(i-1), {\bf u}(w_{2}|s_{i-1})\right)
 \in A_{\epsilon}^{(n)}\right\}$.  Now, $R_{x2}$ uses $s_i$ to find a unique $\hw_{2,i-1} \in S_{s_i} \bigcap
 \mathcal{L}(i-1)$. If there is none or there is more than one, an error is declared.
\end{enumerate}

\subsubsection{Analysis of the Probability of Error}
\label{sec:Pe_analysis_phy_deg}

The achievable rate to $\Rbad$ can be proved using the same technique as in \cite[theorem 1]{CoverG:79}.
For the ease of description assume that $\Rgood$ is connected via an orthogonal channel to $\Rbad$ and let $X'$ denote
the channel input from $\Rgood$ and $Y'$ the corresponding channel output to $\Rbad$. Thus,
$\Rbad$ has combined input $(Y_2,Y')$. The overall transition matrix is given by
\begin{equation}
\label{eqn:Phy-deg-Markov-chain}
        p(y_1,y_2,y'|x,x') = p(y_1,y_2|x)p(y'|x').
\end{equation}
Additionally, we select the transition matrix $p(y'|x')$  and the
input and output alphabets $\mX'$, $\mY'$ such that the capacity of
the orthogonal channel $X' -  Y'$ is $C_{12}$. An example for such a
selection is letting $\mX'$ = $\mY'$ = $\left\{0,1,...,2^{\lceil C_{12} \rceil}-1\right\}$, where
$\lceil \cdot \rceil$ is denotes the ceil function.
Letting  $[a]$ denotes the integer part of the real number $a$, we set the channel transition function to be
\begin{eqnarray*}
    p(Y'|X') = \left\{
        \begin{array}{cl}
            1- \alpha     & ,Y' = X'  \\
            \alpha        & ,Y' = \mod\left(X' + 2^{[C_{12}]},2^{\lceil C_{12} \rceil}\right),
        \end{array}
    \right.
\end{eqnarray*}
with $\alpha$ selected such that $H(Y'|X') = \lceil C_{12} \rceil - C_{12}$.
The capacity of this channel is $C_{12}$ and is achieved by letting $p(x') = \frac{1}{2^{\lceil C_{12} \rceil}}$,
$\forall x' \in \mX'$.
This setup is equivalent to the original setup described in section \ref{sec:generalBCintro}.

Now consider the rate to $\Rbad$. The Markov chain $U - X - (Y_1,Y_2)$ combined with the condition in
(\ref{eqn:Phy-deg-Markov-chain}) implies the following probability distribution function (p.d.f.)
\[
    p(u,y_1,y_2,y',x') = p(y_1,y_2|u)p(y'|x')p(u,x').
\]
Now, applying \cite[theorem 1]{CoverG:79}, with $p(u,x') = p(u)p(x')$, we have that (see also \cite{GuptaKumar:2003})
\begin{eqnarray*}
              R_2 & \le & \min\left\{I(U,X';Y_2,Y'),I(U;Y_1|X') \right\}\\
                  &  =  & \min\left\{I(U,X';Y') + I(U,X';Y_2|Y') , I(U;Y_1) \right\}\\
                  &  =  & \min\left\{I(X';Y') +  I(U;Y'|X') + I(U;Y_2|Y')  \right.\\
                  &     & \qquad \qquad \left.+I(X';Y_2|Y',U) , I(U;Y_1)\right\}\\
                  &  =  & \min\left\{C_{12} + I(U;Y_2), I(U;Y_1)\right\}.
\end{eqnarray*}
Next, consider the rate to $\Rgood$. From the proof of \cite[theorem 1]{CoverG:79} we have that
$\Rgood$ decodes $W_2$. Therefore, $\Rgood$ can now use successive decoding similar to the decoding
at $\Rgood$ in \cite[Ch. 14.6.2]{cover-thomas:it-book}, which imply that the achievable rate to $\Rgood$ is given by
$R_1 \le I(X;Y_1|U)$. Combining both bounds we get the rate constraints of theorem~\ref{thm:converse}.

\subsection{Converse Proof}
\label{sec:converse}

In this section we prove that for $\Perr \rightarrow 0$, the rates
must satisfy the constraints in theorem \ref{thm:converse}. First,
note that for the case of the physically degraded broadcast channel
with cooperating receivers we have the following Markov chain:
\begin{equation}
\label{eqn:markov_chain}
  X^n - Y^n_1 - \big(W_{12}(Y^n_1),Y^n_2\big).
\end{equation}

Considering the definition of the decoders
in~(\ref{eqn:g1_coop}) and~(\ref{eqn:g2_coop}), and the definition
of the probability of error for each of the receivers
in~(\ref{eqn:def_pe1_II}) and~(\ref{eqn:def_pe2_II}), we have from Fano's
inequality (\cite[Ch. 2.11]{cover-thomas:it-book}) that
\begin{eqnarray}
\label{eqn:fano1}
H(W_1|Y_1^n) \!\!
  & \le & \!\!P_{e1}^{(n)}\log_2\left(2^{nR_1}-1\right) + h(P_{e1}^{(n)}) \\
          &  & \triangleq n\delta(P_{e1}^{(n)}),\nonumber\\
\label{eqn:fano2}
\!\!\!\!\!\!\!\!\!H(W_2|Y_2^n,W_{12}(Y_1^n))
  \!\!& \le &\!\! P_{e2}^{(n)}\log_2\left(2^{nR_2}-1\right)  + h(P_{e2}^{(n)})  \\
          &  & \triangleq n\delta(P_{e2}^{(n)}),\nonumber
\end{eqnarray}
where $h(P)$ is the entropy of a Bernoulli RV with parameter $P$.
Note that when
$P_{e1}^{(n)} \rightarrow 0$ then $\delta(P_{e1}^{(n)})
\rightarrow 0$ and when $P_{e2}^{(n)} \rightarrow 0$ then
$\delta(P_{e2}^{(n)}) \rightarrow 0$.

Now, for $\Rgood$ we have that
\begin{eqnarray}
nR_1 & = & H(W_1)  =  I(W_1;Y_1^n) + H(W_1|Y_1^n) \nonumber.
\end{eqnarray}
Applying inequality (\ref{eqn:fano1}), and then proceeding as in
\cite{Gallager:74} we get the bound on $R_1$ as

\[
   nR_1 \le \sum_{k=1}^n I(X_k;Y_{1,k}|U_k) + n\delta(P_{e1}^{(n)}),
\]
where $U_k \triangleq \left(Y_{1,1}, Y_{1,2},...,Y_{1,k-1},W_2\right)$.

For $R_{x2}$ we can write
\begin{eqnarray}
nR_2 & = & H(W_2) \nonumber \\
%  &  =  & H(W_2) - H(W_2|Y_2^n,W_{12}(Y_1^n))  + H(W_2|Y_2^n,W_{12}(Y_1^n))\nonumber\\
  \label{eqn:R2_trns_3}
  & \stackrel{(a)}{\le} & I(W_2;Y_2^n,W_{12}(Y_1^n)) + n\delta(P_{e2}^{(n)})\\
  &  =  & I(W_2;Y_2^n) + I(W_2;W_{12}(Y_1^n)|Y_2^n) + n\delta(P_{e2}^{(n)}),
          \nonumber
\end{eqnarray}
where the inequality in~(a) is due to (\ref{eqn:fano2}).
Proceeding as in \cite{Gallager:74}, we bound $I(W_2;Y_2^n) \le \sum_{k=1}^nI(U_k;Y_{2,k})$. % resulting in
%\[
%    nR_2 \le \sum_{k=1}^nI(U_k;Y_{2,k})  + I(W_2;W_{12}(Y_1^n)|Y_2^n)
%          + n\delta(P_{e2}^{(n)}).
%\]
Next, we bound $I(W_2;W_{12}(Y_1^n)|Y_2^n)$ as follows:
\begin{eqnarray}
\label{eqn:coop_rate_bound}
I(W_{12}(Y_1^n);W_2|Y_2^n) & \le & H(W_{12}(Y_1^n)|Y_2^n)\nonumber\\
                    & \le & H(W_{12}(Y_1^n))\nonumber\\
                    & \le & nC_{12},
\end{eqnarray}
where the first inequality follows from the definition of mutual information, the second
is due to removing the conditioning and the third is due to the admissibility of the
conference. Combining both bounds we get that
\begin{equation}
\label{eqn:R2_rate_sec1}
   nR_2 \le \sum_{k=1}^nI(U_k;Y_{2,k}) + nC_{12} + n\delta(P_{e2}^{(n)}).
\end{equation}

The bound on $R_2$ can be developed in an alternative way.  Begin
with~(\ref{eqn:R2_trns_3}):
\begin{eqnarray}
\!\!\!\!\!\!nR_2
  & \le & I(W_2;Y_2^n,W_{12}(Y_1^n)) + n\delta(P_{e2}^{(n)})\nonumber\\
  & \stackrel{(a)}{\le} & I(W_2;Y_2^n,Y_1^n)+n\delta(P_{e2}^{(n)}) \nonumber\\
    \label{eqn:r2_trans_5}
  &  =  & \sum_{k=1}^n I(W_2; Y_{1,k},Y_{2,k}|Y_1^{k-1},Y_2^{k-1})
          + n\delta(P_{e2}^{(n)}),
\end{eqnarray}
where (a) follows from the fact that
$(W_1,W_2) - (Y_1^n,Y_2^n) - (W_{12},Y_2^n)$ is a Markov relation and
from the data processing inequality. Next, we can write
\begin{eqnarray}
&  &I(W_2; Y_{1,k}, Y_{2,k} |Y_1^{k-1},Y_2^{k-1})\nonumber\\
%  &  =  & H(Y_{1,k},Y_{2,k}|Y_1^{k-1},Y_2^{k-1})  - H(Y_{1,k},Y_{2,k}|Y_1^{k-1},Y_2^{k-1},W_2) \nonumber \\
%  &  =  & H(Y_{1,k}|Y_1^{k-1},Y_2^{k-1}) + H(Y_{2,k}|Y_{1,k},Y_1^{k-1},Y_2^{k-1}) \nonumber \\
%  &     & \quad - H(Y_{1,k}|Y_1^{k-1},Y_2^{k-1},W_2)  - H(Y_{2,k}|Y_{1,k},Y_1^{k-1},Y_2^{k-1},W_2)\nonumber \\
  &  & \qquad \stackrel{(a)}{ = }I(W_2; Y_{1,k} |Y_1^{k-1},Y_2^{k-1})\nonumber \\
  &  & \qquad = H(Y_{1,k}|Y_1^{k-1},Y_2^{k-1})
          - H(Y_{1,k}|Y_1^{k-1},Y_2^{k-1},W_2) \nonumber\\
%          \label{eqn:trans_7}
  &  & \qquad \stackrel{(b)}{\le} H(Y_{1,k}) - H(Y_{1,k}|Y_1^{k-1},Y_2^{k-1},W_2)\nonumber\\
  &  & \qquad \stackrel{(c)}{=} H(Y_{1,k}) - H(Y_{1,k}|Y_1^{k-1},W_2)\nonumber\\
  &  & \qquad = I(Y_{1,k};Y_1^{k-1},W_2) \nonumber \\
  &  & \qquad = I(Y_{1,k};U_k),
\end{eqnarray}
where the equality in~(a) is due to the physical degradedness and memorylessness of the channel,
(b) is due to removing the conditioning, and~(c) is because the Markov chain makes
$Y_{1,k}$ independent of $Y_2^{k-1}$ given $Y_1^{k-1}$.
Plugging this into~(\ref{eqn:r2_trans_5}), we get a second bound
on $R_2$:
\[
   nR_2 \le \sum_{k=1}^n I(U_k;Y_{1,k}) + n\delta(P_{e2}^{(n)}).
\]

Collecting the three bounds we have:
\begin{eqnarray}
\label{eqn:sum_covrs_1}
  R_1 & \le & \frac 1 n\sum_{k=1}^n I(X_k;Y_{1,k}|U_k)+\delta(P_{e1}^{(n)}), \\
  R_2 & \le & \frac 1 n\sum_{k=1}^nI(U_k;Y_{2,k}) + C_{12}
              + \delta(P_{e2}^{(n)}), \\
\label{eqn:sum_covrs_2}
  R_2 & \le & \frac 1 n\sum_{k=1}^n I(U_k;Y_{1,k}) + \delta(P_{e2}^{(n)}).
%  R_1 + R_2 & \le & \frac 1 n\sum_{k=1}^nI(X_k;Y_{1,k})
%                    + (\delta(P_{e1}^{(n)}) + \delta(P_{e2}^{(n)}))
\end{eqnarray}
Using the standard time-sharing argument as in~\cite[Ch.\ 14.3]{cover-thomas:it-book}, we can write the averages
in~(\ref{eqn:sum_covrs_1}) - (\ref{eqn:sum_covrs_2}) by introducing an appropriate time sharing variable, with cardinality
upper bounded by $4$. Therefore, if $P_{e1}^{(n)} \rightarrow 0$ and $P_{e2}^{(n)} \rightarrow 0$ as
$n \rightarrow \infty$, the convex hull of this region can be shown to be equivalent
to the convex hull of the region defined by
\begin{eqnarray}
    \label{eqn:converse_1r}
    R_1    & \le & I(X;Y_1|U), \\
    \label{eqn:converse_2r}
    R_2    & \le & I(U;Y_2) + C_{12}, \\
    \label{eqn:converse_three_r}
    R_2    & \le & I(U;Y_1).
\end{eqnarray}

Finally, the bound on the cardinality of $\mU$ follows from the same arguments as in the
converse  for the non-cooperative case in~\cite{Gallager:74}. Note however, that
$|| \mY_2 ||$ is absent from the minimization on the cardinality (cf.
equation (\ref{eqn:U_bound_no_coop_degradd}) for the non-cooperative case).
The reason is that even when $|| \mY_2 || = 1$, information to $\Rbad$ (represented
by the random variable $U$), can be sent through the conference link between the two receivers. \tend

\subsection{Discussion}
\begin{figure}[h]
    \centering
    \scalebox{0.62}{\includegraphics{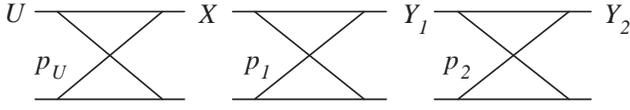}}
    \caption{\small The physically degraded BSBC. $p_U$, $p_1$ and $p_2$ are the transition probabilities
        at the left, middle and right segments respectively.}
    \label{fig:broadcast-degraded_channel}
\end{figure}

\begin{figure}[ht]
    \centering
    \includegraphics[width = 0.42\textwidth]{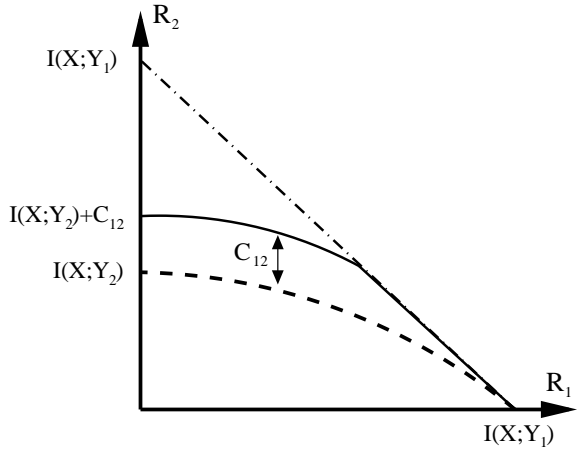}
    \caption{\small The capacity region for the physically degraded BSBC. Top, middle and bottom lines
        correspond to maximum possible cooperation, partial cooperation and no-cooperation scenarios respectively.}
    \label{fig:broadcast-degraded_example}
\end{figure}

To illustrate the implications of theorem \ref{thm:converse}, consider the physically degraded binary symmetric
broadcast channel (BSBC) depicted in figure \ref{fig:broadcast-degraded_channel}.
For this channel, theorem \ref{thm:converse} implies that $||\mU|| = 2$. Due to the symmetry of the channel,
the probability distribution of $U$ which maximizes the rates, is a symmetric binary distribution,
$\Pr(U=0) = \Pr(U=1) = \frac{1}{2}$. The resulting capacity region for this case is depicted in
figure \ref{fig:broadcast-degraded_example} for the case where $R_0 = 0$.
In the figure, the bottom line (dash) is the non-cooperative capacity region, and the top line (dash-dot)
is the maximum possible sum rate, which requires that $C_{12} \ge h(p_{12}) - h(p_1)$, where
\begin{eqnarray*}
    h(p)   & = & -p\log_2(p) - (1-p)\log_2(1-p),\\
    p_{12} & = & p_1(1-p_2) + p_2(1-p_1).
\end{eqnarray*}
This maximum sum-rate of $I(X;Y_1)$ is obtained by summing the rate to $\Rgood$ given by (\ref{eqn:converse_1r}) and
the maximum possible rate for $\Rbad$ given by (\ref{eqn:converse_three_r}), and using the
Markov chain relation $U - X - Y_1$.
The middle line (solid) is the capacity region for the partial cooperation case where $0 < C_{12} < h(p_{12}) - h(p_1)$.

As can be seen from this example, the capacity region derived in this section is  strictly
larger than the capacity region for the non-cooperation case.
Indeed, summing the constraints on $R_0$, $R_1$ and $R_2$ without
cooperation (equations (\ref{eqn:Degraded_NoCoope_1}),
(\ref{eqn:Degraded_NoCoope_2})), results in a maximum achievable
sum-rate of
\begin{equation}
\label{eqnLsum_rate_deg_no_coop}
   R_0 + R_1 + R_2 \le I(X;Y_1) - (I(U;Y_1) - I(U;Y_2)),
\end{equation}
where the second term is always positive due to the Markov chain
$U-X-Y_1-Y_2$ (assuming the degrading channel
is non-invertible\footnote{It can be shown that $I(U;Y_1) - I(U;Y_2) = 0$ for the degraded channel setup
implies that if $R_0 + R_2 > 0$ then $H(Y_1|Y_2) = 0$, i.e. the channel from $\Rgood$ to $\Rbad$ is invertible.
Under these circumstances, this setup can be replaced by an equivalent setup in which
both receivers get $Y_1$, but such a degenerate setup is not interesting.}).
In this setup, the maximum possible sum-rate,
$I(X;Y_1)$, is achieved only when $U$ is a constant, and thus no
information is sent to $R_{x2}$.  When $R_0 + R_2 > 0$, because of the relationship
$R_0 + R_2 \le I(U;Y_2) < I(U;Y_1)$, we cannot achieve the
maximum sum-rate of $I(X;Y_1)$ to
$R_{x1}$.
%The loss in sum-rate is exactly this positive difference, $I(U;Y_1) - I(U;Y_2)$.
However, summing~(\ref{eqn:converse_2r})
or~(\ref{eqn:converse_three_r}) with~(\ref{eqn:converse_1r}),
results in a maximum achievable sum-rate with cooperating receivers of
\begin{eqnarray}
&  & R_0 + R_1 + R_2  \le  I(X;Y_1)\nonumber\\
        &   & \phantom{xxxxxxx}+\min\left\{0,C_{12}-(I(U;Y_1)-I(U;Y_2))\right\}.
\end{eqnarray}
Comparing this to non-cooperative sum-rate given by (\ref{eqnLsum_rate_deg_no_coop}),
it is clear that cooperation allows a net increase in the sum-rate, by at most $C_{12}$.

%%%%%%%%%%%%%%%%%%%%%%%%%%%%%%%%%%%%%%%%%%%%%%%%%%%%%%%%%%%%%%%%%%%%%%%%%%%%%%%%%%%%%%%%%%%%%%%%%%%%%%%%%%%%%
%%%%%%%%%%%%%%%%%%%%%%%%%%%%%%%%%%%%%%%%%%%%%%%%%%%%%%%%%%%%%%%%%%%%%%%%%%%%%%%%%%%%%%%%%%%%%%%%%%%%%%%%%%%%%
%%%%%%%%%%%%%%%%%%%%%%%%%%%%%%%%%%%%%%%%%%%%%%%%%%%%%%%%%%%%%%%%%%%%%%%%%%%%%%%%%%%%%%%%%%%%%%%%%%%%%%%%%%%%%

\section{Achievable Rates for the General Broadcast Channel with Cooperating Receivers}
\label{sec:generalBC}

For the classic general BC scenario, the best achievability result was derived
by Marton in \cite{Marton:79}. This result states that for the general BC,
any rate pair $(R_1,R_2)$ satisfying
\begin{eqnarray}
\label{eqn:marton_R1}
    R_1 & \le & I(U;Y_1),\\
\label{eqn:Marton_R2}
    R_2 & \le & I(V;Y_2),\\
\label{eqn:Marton_SumRate}
    R_1 + R_2 & \le & I(U;Y_1) + I(V;Y_2) - I(U;V),
\end{eqnarray}
for some joint distribution $p(u,v,x,y_1,y_2)=p(u,v,x)p(y_1,y_2|x)$,
is achievable.

We note that Marton's largest region contains three auxiliary RVs,
$(W,U,V)$, where $W$ represents information decoded by both receivers.  Here we
use a simplified version, where $W$ is set to a constant.

We now consider cooperation between the receivers. We begin with a statement of the theorem:
\begin{theorem}
\label{thm:achieve}
\it
Let $\left(\mathcal{X}, p(y_1,y_2|x), \mathcal{Y}_1 \times \mathcal{Y}_2\right)$ be any discrete memoryless
broadcast channel, with cooperating
receivers having noiseless conference links of finite capacities $C_{12}$ and $C_{21}$, as defined in Section~\ref{sec:defs}.
Then, for sending independent information, any rate pair $\left(R_1,R_2 \right)$ satisfying
\begin{eqnarray*}
  R_1       & \le & R(U),\\
  R_2       & \le & R(V),\\
  R_1 + R_2 & \le & R(U) + R(V) - I(U;V),
\end{eqnarray*}
subject to,
\begin{eqnarray}
    \label{eqn:C21_rate conbtraint_Rev}
    C_{21} & \ge & I(\hat{U};Y_2) - I(\hU;Y_1),\\
    \label{eqn:C12_rate conbtraint_Rev}
    C_{12} & \ge & I(\hat{V};Y_1) - I(\hV;Y_2),
\end{eqnarray}
where,
\begin{eqnarray}
    \label{eqn:def_RU}
    R(U) & = & I(U;Y_1,\hU),\\
    \label{eqn:def_RV}
    R(V) & = & I(V;Y_2,\hV),
\end{eqnarray}
for some joint distribution $p(u,v,x,y_1,y_2,\hat{u},\hat{v}) =
p(u,v,x)p(y_1,y_2|x)p(\hat{u}|y_2)p(\hat{v}|y_1)$, is achievable,
with $u \in \mU, v \in \mV, \hu \in \mhU, \hv \in \mhV$, $||\mhU||
\le ||\mY_2||+1$ and $||\mhV|| \le ||\mY_1||+1$.
\end{theorem}
%\smallskip
In the next subsections we provide the proof of this theorem.

\subsection{Overview of Coding Strategy}
As in the achievability part of theorem \ref{thm:converse}, the proposed code is a hybrid broadcast-relay code.
Here, we combine the relay code construction of \cite[theorem 6]{CoverG:79} and the broadcast code construction of
\cite{ElGamalM:81}. The fact that in these two theorems the channel encoding and the relay operation
are performed independently, allows to easily combine them into a hybrid coding scheme.
The encoder generates broadcast codewords, each selected from a codebook constructed similarly to the
construction of \cite{ElGamalM:81}. This codebook splits the rate between
the two users. Next, each relay ($R_{x1}$ acts as a relay for
$R_{x2}$ and vice-versa) generates its codebook according to the
construction of \cite[theorem 6]{CoverG:79}. In the decoding
step, using the received signal ($Y_1^n$ at $\Rgood$ and $Y_2^n$
at $\Rbad$), each receiver generates a list of the possible
transmitted relay messages and uses the conference message from the next time interval to
resolve for the relay massage. Then, each receiver uses the decoded relay
message and its received channel output to decode its own message.
%In our decoding scheme,
%after receiving the conference message the receiver decodes the source message by looking for a source
%sequence ($U^n$ for $\Rgood$ and $V^n$ for $\Rbad$) such that it is jointly typical with its channel
%input and with at least one of the relay sequences in the set received from the other receiver.
%We show that this results in an increased feasible region, see comment \ref{cmt:increased_region}.

\subsection{Encoding at the Transmitter}
\begin{enumerate}
\item Let $\epsilon > 0$ and $n \ge 1$ be given. Fix $p(u,v,x)$, $p(\hu|y_2)$ and $p(\hv|y_1)$, and let
      $\delta > 0$ be a positive number, whose selection is described in the next item.
      Let $\stypd(U)$ denote the set of strongly typical i.i.d. sequences of length $n$, $\uvec \in \mU^n$,
      as defined in \cite[Ch. 13.6]{cover-thomas:it-book}.
      Let $\stypd(V)$ denote the set of strongly typical i.i.d. sequences of length $n$, $\vvec \in \mV^n$.
      Let $S_{[U]\delta}^{(n)}$ denote the set of all sequences $\uvec \in \stypd(U)$, such that $\stypd(V|\uvec)$
      is nonempty as defined in \cite[corollary 5.11]{YeungBook},
      and similarly define $S_{[V]\delta}^{(n)}$ for the sequences $\vvec \in \stypd(V)$.

\item  Select $2^{n\left(R(U) - \epsilon\right)}$ strongly typical sequences $\uvec$ in an i.i.d. manner,
according to the probability
\begin{eqnarray*}
    p\left(\uvec\right) = \left\{
        \begin{array}{cl}
            \frac{1}{||S_{[U]\delta}^{(n)}||} &, \uvec \in S_{[U]\delta}^{(n)}\\
            0 &, \mbox{otherwise}.
        \end{array}
    \right.
\end{eqnarray*}
%%$p(\uvec) = \prod_{i=1}^n p(u_i)$.
%where $||A||$ denotes the cardinality of set $A$.
Label these sequences by $\uvec(k), \; k\in\Big[1, 2^{n\left(R(U)-\epsilon\right)}\Big]$.
%\footnote{
%note that $\Iu = H(U)-H\left(U|Y_1,\hat{U}\right) \le H(U)$, so we can find the required number of sequences}.
Select $2^{n\left(R(V) - \epsilon\right)}$ strongly typical sequences $\vvec$ in an i.i.d. manner,
according to the probability
\begin{eqnarray*}
    p\left(\vvec\right) = \left\{
        \begin{array}{cl}
            \frac{1}{||S_{[V]\delta}^{(n)}||} &, \vvec \in S_{[V]\delta}^{(n)}\\
            0 &, \mbox{otherwise}.
        \end{array}
    \right.
\end{eqnarray*}
%$p(\vvec) = \prod_{i=1}^n p(v_i)$.
Label these sequences by $\vvec(l), \; l\in\left[1,2^{n\left(R(V)-\epsilon\right)}\right]$.
Note that from \cite[corollary 5.11]{YeungBook} we have that $||S_{[U]\delta}^{(n)}|| \ge (1-\delta)2^{n(H(U) - \psi)}$,
where $\psi \rightarrow 0$ as $n \rightarrow \infty$ and $\delta \rightarrow 0$, so for any $\eps > 0$ we can always
find $0 < \delta \le \eps$ such that for $n$ large enough we obtain $||S_{[U]\delta}^{(n)}|| > 2^{n\left(\Iu - \epsilon\right)}$
and $||S_{[V]\delta}^{(n)}|| > 2^{n\left(\Iv - \epsilon\right)}$.
%and $R(U) \le \Iu$. Similarly this restriction
%does not impede on the selection of the $\vvec$ sequences.

\item Define the cells $B_i = \left[(i-1)2^{n\left(R(U) - R_1 - \epsilon\right)}+1,
    i 2^{n\left(R(U) - R_1 - \epsilon\right)}  \right]$, $i \in \left[1,2^{nR_1}\right]$.
    This is a partition of the $\uvec$ sequences into $2^{nR_1}$ sets. Define the cells \\
    $C_j = \left[(j-1)2^{n\left(R(V) - R_2 - \epsilon\right)}+1,
    j 2^{n\left(R(V) - R_2 - \epsilon\right)}  \right]$, $j \in \left[1,2^{nR_2}\right]$, which form a
    partition of the $\vvec$ sequences into $2^{nR_2}$ sets.

\item For every pair of integers $(w_1,w_2) \in \left[1,2^{nR_1} \right]\times \left[1,2^{nR_2} \right]$,
define the set
%\begin{eqnarray*}
$\mD_{w_1,w_2}  =  \Big\{ \left(\uvec(k), \vvec(l)\right): k \in B_{w_1}, l\in C_{w_2},  %\\
                  \left. \left( \uvec(k), \vvec(l)\right) \in \styp(U,V) \right\}.$
%\end{eqnarray*}
% containing the strongly joint typical pairs $(\uvec,\vvec)$ in
%$B_{w_1} \times C_{w_2}$.
Here, $\styp (U,V)$ denotes the strongly typical set for the random
variables $U$ and $V$ as defined in \cite[Ch. 13.6]{cover-thomas:it-book}.
In the following we may omit the random variables when referring to the strongly typical set,
 when these variables are clear from the context.
We now have the following (slightly modified) lemma from \cite{ElGamalM:81}:
\begin{lemma}
\label{lemma:lemma_1}
{\it
For any 2-D cell $B_i \times C_j$, $\epsilon > 0$, and $n$ large enough, we have that
$\mbox{Pr}\left(||\mD_{ij}|| = 0 \right) \le \epsilon$,
provided that
\begin{eqnarray}
\label{eqn:lemma1}
  \!\!\!\!\!\!\!\!\!\! R_1 + R_2  & < &  R(U) + R(V) -I(U;V) - 2\epsilon - \epsilon_1,
\end{eqnarray}
where $\epsilon_1 \rightarrow 0$ as $\epsilon \rightarrow 0$ and $n \rightarrow \infty$.}
\end{lemma}
\begin{proof}
    The proof of this lemma is obtained by direct application of the
    technique used to prove \cite[Lemma in pg.\ 121]{ElGamalM:81}, and therefore will not be repeated here.
\end{proof}

\item For each message pair $(w_1,w_2)$, select one pair $(\uvec(k_{w_1,w_2}), \vvec(l_{w_1,w_2}))\in \mD_{w_1,w_2}$. For
each of the selected pairs (one pair for each message pair), generate a codeword according to
$\xvec(w_1,w_2) \sim \prod_{i=1}^n p\left(x_i | u_i(k_{w_1,w_2}), v_i(l_{w_1,w_2})\right)$.

\item To transmit the message pair $(w_1,w_2)$ the transmitter outputs $\xvec(w_1,w_2)$.
\end{enumerate}

\subsection{Encoding the Relay Messages}
Consider first the relay encoding at $R_{x2}$, which acts as a relay for $R_{x1}$.
\begin{enumerate}
\item $R_{x2}$-relay has a set of $2^{nC_{21}}$ relay messages indexed by $s' \in \left[1,2^{nC_{21}}\right]$.
For each index $s'$, generate $2^{nR'}$ i.i.d. sequences $\huvec$, each with probability $p(\huvec) = \prod_{i=1}^n p(\hu_i)$, \\
 $p(\hu) = \sum_{\mX, \mY_1, \mY_2} p(\hu|y_2)p(y_1,y_2|x)p(x)$, and
$p(x) = \sum_{\mU, \mV} p(u,v,x)$. Label these codewords $\huvec(z'|s')$, $s' \in \left[1,2^{nC_{21}}\right]$,
$z' \in [1,2^{nR'} ]$.

\item Randomly and uniformly partition the message set $[1,2^{nR'} ]$ into $2^{nC_{21}}$ sets $S'_{s'}$,
$s' \in \left[1,2^{nC_{21}}\right]$.

\item \underline{Encoding}: Assume that after receiving $\yvec_2(i-1)$
%the channel output sequence for the $(i-1)$'th codeword %out of the $B-1$ codewords,
we have
at $R_{x2}$ that $\left( \huvec(z_{i-1}'|s_{i-1}'), \yvec_2(i-1) \right) \in \styp$, and that $z_{i-1}' \in S'_{s_i'}$
($s_{i-1}'$ is known from the previous transmission of $z_{i-2}'$).
Then, at the $i$'th transmission interval the relay transmits the index $s_i'$ to $R_{x1}$.
\end{enumerate}
Relay encoding at $R_{x1}$ is performed in a symmetric manner to the relay encoding at
 $R_{x2}$. The corresponding variables for $\Rgood$ are $S''_{s''}$ and $\hvvec(z''|s'')$,
 $s'' \in \left[1,2^{nC_{12}}\right]$, $z'' \in [1,2^{nR''}]$.

\subsection{Decoding the Relay Messages at the Relays}
\label{sec:dec_rly_2}
Consider decoding the relay message at $R_{x2}$.
The relay decoder at $R_{x2}$ uses its channel input $\yvec_2(i)$, and its previously decoded $s_i'$
to generate the relay message $z_i'$ as follows:
upon receiving $\yvec_2(i)$, the relay $R_{x2}$ decides that the message $z_i'$ was received at time $i$ if
$\left(\huvec\left(z_i'|s_i'\right), \yvec_2(i)\right) \in \styp$.
Following the argument in \cite[theorem 6]{CoverG:79} (see
also the proof in \cite[Ch. 13.6]{cover-thomas:it-book}), there exists such $z_i'$ with
probability that is arbitrarily close to one  as long as
\begin{equation}
\label{eqn:dec_ar_rly}
    R' \ge I\left(\hU;Y_2\right),
\end{equation}
and $n$ is sufficiently large.
Relay decoding at $R_{x1}$ is done in a symmetric manner to the relay decoding at $R_{x2}$.

\subsection{Decoding at the Receivers}
\label{sec:dec_rx_1} We first find the rate constraint for decoding
at $\Rgood$. $R_{x1}$ decodes its message $w_{1,i-1}$ based on its
channel input $\yvec_1(i-1)$ and the relay indices $s_i'$ and $s_{i-1}'$:
\begin{enumerate}
\item From knowledge of $s_{i-1}'$ and $\yvec_1(i-1)$, $R_{x1}$ calculates the set $\mL_1(i-1)$ such that
    \begin{eqnarray*}
        \mL_1(i-1)  & = &  \Big\{z' \in [1,2^{nR'} ]: \\
                &  & \qquad\left(\huvec\left(z'|s_{i-1}'\right),\yvec_1(i-1)\right) \in \styp \Big\}.
    \end{eqnarray*}
\item At the time interval of the $i$'th codeword, $R_{x1}$ receives the relayed
    $s_i'$. Since $s_i'$ is selected from a set of $2^{nC_{21}}$
    possible messages, it can be transmitted over the noiseless conference
    link without error.

\item $R_{x1}$ now chooses $\hz_{i-1}'$ as the relay message at time $i-1$ if and only if there exists a unique
     $\hz_{i-1}' \in S'_{s_i'} \bigcap \mL_1(i-1)$. Again, following the reasoning in \cite[theorem 6]{CoverG:79},
     %(see also the derivation leading to equation (\ref{eqn:R2_second_bound_K}) in section \ref{sec:pe_bad_degraded})
     this can be done with an arbitrarily small probability of error as long as
     \begin{equation}
     \label{eqn:upper_bound_Rprime}
         R' \le I(\hU;Y_1) + C_{21},
     \end{equation}
     and $n$ is large enough.
     Combining this with inequality (\ref{eqn:dec_ar_rly}) we get the constraint on the relay information rate:
     \begin{equation}
     \label{eqn:C_two_one_constr}
         C_{21} \ge I(\hU;Y_2) - I(\hU;Y_1).
     \end{equation}
     This expression is similar to the Wyner-Ziv expression for the rate required to transmit $Y_2$ to receiver $\Rgood$
      up to a given distortion, determined by $p(\hu|y_2)$ and a decoder.
      Here the performance of the decoder are implied in the mutual information $I(U;Y_1,\hU)$.
      The compressed $Y_2^n$ is then used by $\Rgood$ to assist in decoding $W_1$.

 \item Lastly, $R_{x1}$ decodes $w_{1,i-1}$ (or, equivalently $\uvec(k_{w_{1,i-1},w_{2,i-1}})$) by choosing
     $\uvec(k_{\hw_{1,i-1},\hw_{2,i-1}})$
     such that
     $\left(\uvec(k_{\hw_{1,i-1},\hw_{2,i-1}}), \yvec_1(i-1), \huvec\left(\hz_{i-1}'|s_{i-1}'\right)  \right) \in \styp$. From
     the point-to-point channel coding theorem (see \cite{ElGamalM:81}) we have that
     $\hat{w}_{1,i-1} = w_{1,i-1}$ with probability that is arbitrarily close to one, as long as $z_{i-1}'$ was correctly
     decoded at $\Rgood$ and
     \begin{equation}
         R_1 \le R(U) \triangleq \Iu,
     \end{equation}
     for sufficiently large $n$.
     Combining this with equation (\ref{eqn:C_two_one_constr}) yields the rate constraint on $R_1$:
     \begin{eqnarray}
     \label{eqn:rate_r1_eq1}
         R_1 & \le & R(U), \\
     \label{eqn:rate_r1_eq2}
         &   & \mbox{as long as  }C_{21} \ge I(\hU;Y_2) - I(\hU;Y_1).
     \end{eqnarray}
\end{enumerate}

 Using symmetric arguments to those
 presented for decoding at $\Rgood$ we find the rate constraint for $R_{x2}$ to be
 \begin{eqnarray}
 \label{eqn:rate_r2_eq1}
      R_2 & \le & R(V), \\
 \label{eqn:rate_r2_eq2}
       &   & \mbox{as long as  }C_{12} \ge I(\hV;Y_1) - I(\hV;Y_2).
 \end{eqnarray}

Combining equations (\ref{eqn:lemma1}),
(\ref{eqn:rate_r1_eq1}), (\ref{eqn:rate_r1_eq2}), (\ref{eqn:rate_r2_eq1}) and (\ref{eqn:rate_r2_eq2}), gives the
conditions in theorem \ref{thm:achieve}.

\subsection{Error Events}
\label{sec:error_event_two_indep}
In the scheme described above we have to account for the following error events for decoding $(w_{1,i-1},w_{2,i-1})$:
\begin{enumerate}
\item Encoding at the transmitter fails: \\
$E_{D,i} = \left\{||\mD_{w_{1,i-1},w_{2,i-1}}|| = 0 \right\}$.

\item Joint typicality decoding fails:\\
    $E_{0,i} = \Big\{\big( \uvec(k_{w_{1,i-1},w_{2,i-1}}), \vvec(l_{w_{1,i-1},w_{2,i-1}}),$\phantom{xxxxxxxx}\\
     \phantom{xxxxx}$\xvec(w_{1,i-1},w_{2,i-1}),\yvec_1(i-1), \yvec_2(i-1)
    \big) \notin  \styp \Big\}$.
\item Decoding at the relays fails: $E_{1,i} = E_{11,i} \bigcup E_{12,i}$, \\
    $E_{11,i} = \Big\{\nexists z' \in [1,2^{nR'}] \; \mbox{s.t.} $ \phantom{xxxxxxxxxxxxxxxx}\\
     \phantom{xxxxxxxxxxx}   $\; \left(\huvec\left(z'|s_{i-1}'\right), \yvec_2(i-1)  \right) \in \styp \Big\}$,\\
    $E_{12,i} = \Big\{\nexists z'' \in [1,2^{nR''}] \; \mbox{s.t.}$ \phantom{xxxxxxxxxxxxxxxx}\\
    \phantom{xxxxxxxxxxx}   $\; \left(\hvvec\left(z''|s_{i-1}''\right), \yvec_1(i-1)  \right) \in \styp \Big\}$.

%\item Decoding at the receivers fails: $E_{2,i} = E_{21,i} \bigcup E_{22,i}$, where
%    $E_{21,i} = E_{20,i}'\bigcup E_{21,i}' \bigcup E_{21,i}''$ and $E_{22,i} = E_{20,i}'' \bigcup E_{22,i}' \bigcup E_{22,i}''$,\\
%    $E_{20,i}' = \left\{\big(\uvec(k_{w_{1,i-1},w_{2,i-1}}),\yvec_1(i-1),\huvec(z_{i-1}'|s_{i-1}')\big)\notin\styp\right\}$,\\
%    $E_{20,i}'' = \left\{\big(\vvec(l_{w_{1,i-1},w_{2,i-1}}),\yvec_2(i-1),\hvvec(z_{i-1}''|s_{i-1}'')\big)\notin\styp\right\}$,\\
%    $E_{21,i}' = \left\{ \nexists z' \in S'_{s_i'} \mbox{ s.t. } \big(\uvec(k_{w_{1,i-1},w_{2,i-1}}),\yvec_1(i-1),\huvec(z'|s_{i-1}')\big)\in\styp\right\}$,\\
%    $E_{21,i}''= \left\{ \exists z' \in S'_{s_i'}, \exists k_{w_1,w_2} \in \mL_1(i-1), w_1 \ne w_{1,i-1} \mbox{ s.t. }
%         \big(\uvec(k_{w_1,w_2}),\yvec_1(i-1),\huvec(z'|s_{i-1}')\big)\in\styp\right\}$,\\
%    $E_{22,i}' = \left\{ \nexists z'' \in S''_{s_i''} \mbox{ s.t. } \big(\vvec(l_{w_{1,i-1},w_{2,i-1}}),\yvec_2(i-1),\hvvec(z''|s_{i-1}'')\big)\in\styp\right\}$,\\
%    $E_{22,i}''= \left\{ \exists z'' \in S''_{s_i''}, \exists l_{w_1,w_2} \in \mL_2(i-1), w_2 \ne w_{2,i-1} \mbox{ s.t. }
%         \big(\vvec(l_{w_1,w_2}),\yvec_2(i-1),\hvvec(z''|s_{i-1}'')\big) \in\styp\right\}$,\\
%    and $\mL_2(i-1) = \left\{l_{w_1,w_2} \in \left[1,2^{n(R(V)-\eps)}\right]: (w_1,w_2) \in \mW_1 \times \mW_2, \left(\vvec(l_{w_1,w_2}),\yvec_2(i-1)\right) \in \styp \right\}$.

%We therefore replace $E_{21,i}$ and $E_{22,i}$
% of section \ref{sec:error_event_two_indep} with the following events
%\begin{enumerate}
\item Decoding the relay message at the receivers fails: $E_{2,i} = E_{21,i} \bigcup E_{22,i}$, where
    $E_{21,i} = E_{21,i}' \bigcup E_{21,i}''$ and $E_{22,i} = E_{22,i}' \bigcup E_{22,i}''$,\\
    $E_{21,i}' = \left\{ z_{i-1}' \notin S'_{s_i'} \bigcap \mL_1(i-1) \right\}$,\\
    $E_{21,i}''= \left\{ \exists \tilde{z}' \ne z_{i-1}' \; \mbox{s.t.} \;  \tilde{z}'
    \in S'_{s_i'} \bigcap \mL_1(i-1)\right\}$,\\
    $E_{22,i}' = \left\{ z_{i-1}'' \notin S''_{s_i''} \bigcap \mL_2(i-1) \right\}$,\\
    $E_{22,i}''= \left\{ \exists \tilde{z}'' \ne z_{i-1}'' \; \mbox{s.t.} \;  \tilde{z}''
    \in S''_{s_i''} \bigcap \mL_2(i-1)\right\}$,\\
    $\mL_2(i-1) \triangleq \Big\{z''\in [1,2^{nR''}]:$\phantom{xxxxxxxxxxxxxxxx}\\
    \phantom{xxxxxxxxxxxxx} $\left(\hvvec\left(z''|s_{i-1}''\right),\yvec_2(i-1)\right) \in \styp \Big\}$.

\item Final decoding at the receivers fails:\\
    $E_{3,i} = E_{31,i} \bigcup E_{32,i}$, where,\\
    $E_{31,i} = \Big\{\big(\uvec(k_{w_{1,i-1}, w_{2,i-1}}),\yvec_1(i-1),$\phantom{xxxxxxxxx}
    $\huvec(z_{i-1}'|s_{i-1}')\big)\notin\styp  \Big\}$
    $\!\!\!\bigcup\Big\{ \exists w_1  \ne w_{1,i-1} \; \mbox{s.t.} $
    $\phantom{XXX}\left(\uvec(k_{w_1, w_2}) ,\yvec_1(i-1),
    \huvec(z_{i-1}'|s_{i-1}')  \right) \in \styp\Big\}$, \\
    $E_{32,i} = \Big\{\big(\vvec(l_{w_{1,i-1}, w_{2,i-1}}),\yvec_2(i-1),$\phantom{xxxxxxxxx}
    $\hvvec(z_{i-1}''|s_{i-1}'')\big)\notin\styp  \Big\}$
    $\!\!\! \bigcup\Big\{ \exists w_2 \ne w_{2,i-1}\; \mbox{s.t.}$
    $\phantom{XXX}\left(\vvec(l_{w_1,w_2}) ,\yvec_2(i-1), \hvvec(z_{i-1}''|s_{i-1}'')  \right) \in \styp\Big\}$.
%\end{enumerate}

\end{enumerate}
We now bound the probability of the error events at time $i$. Note that at time $i$ both $\Rgood$ and $\Rbad$ share the same
$s_{i-1}'$ and $s_{i-1}''$  irrespective whether the decoding at the relays was correct at time $i-1$. Hence, a
decoding error at time $i-1$ does not affect the decoding at time $i$. Now,
from lemma \ref{lemma:lemma_1} it follows that by taking $n$ large enough the probability
of $E_{D,i}$ can be made arbitrarily small, as long as (\ref{eqn:lemma1}) is satisfied. Additionally, by taking $n$
large enough, the probability $\Pr(E_{0,i}\bigcap E_{D,i}^c)$ can be made arbitrarily small by the properties
of strongly typical sequences, see \cite[lemma 13.6.2]{cover-thomas:it-book}.
The probability $\Pr(E_{1,i})$ can be made arbitrarily small as long as
(\ref{eqn:rate_r1_eq2}) and (\ref{eqn:rate_r2_eq2}) are satisfied, as explained is section \ref{sec:dec_rly_2}.
Next, the Markov lemma \cite[lemma 4.2]{BetgerLecNotes} and the Markov chains $Y_1 - Y_2 - \hU$ and
$Y_2 - Y_1 - \hV$, imply that $\Pr(E_{21,i}' \bigcap E_{1,i}^c\bigcap E_{0,i}^c)$ and
$\Pr(E_{22,i}' \bigcap E_{1,i}^c \bigcap E_{0,i}^c)$ can be
made arbitrarily small by taking $n$ large enough, and
$\Pr(E_{21,i}'' \bigcap E_{i,1}^c)$ and $\Pr(E_{22,i}'' \bigcap E_{i,1}^c)$ can be made arbitrarily small
by taking $n$ large enough as long as \eqref{eqn:rate_r1_eq2} and \eqref{eqn:rate_r2_eq2} are satisfied.
Finally, $\Pr(E_{31,i} \bigcap E_{2,i}^c \bigcap E_{1,i}^c \bigcap E_{0,i}^c \bigcap E_{D,i}^c)$ and
$\Pr(E_{32,i} \bigcap E_{2,i}^c \bigcap E_{1,i}^c \bigcap E_{0,i}^c \bigcap E_{D,i}^c)$ can be made arbitrarily
small by taking $n$ large enough by the Markov lemma and the chains $U,Y_1 - Y_2 - \hU$ and $V,Y_2 - Y_1 - \hV$,
and as long as \eqref{eqn:rate_r1_eq1} and \eqref{eqn:rate_r2_eq1} are satisfied.

%$\Pr(E_{20,i}' \bigcup E_{20,i}'' \bigcap E_{1,i}^c \bigcap E_{0,i}^c)$ can be made
%arbitrarily small by taking $n$ large enough.
%We also have that $\Pr(E_{21,i}'\bigcap E_{20,i}^{'c} \bigcap E_{1,i}^c \bigcap E_{0,i}^c)$
%$ = \Pr(E_{22,i}'\bigcap E_{20,i}^{''c} \bigcap E_{1,i}^c \bigcap E_{0,i}^c) = 0$ since $E_{1,i}^c$ implies
%successful decoding at the relay at time $i-1$ and by construction $z_{i-1}' \in S'_{s_i'}$ and
%$z_{i-1}'' \in S''_{s''_i}$. Finally, appendix \ref{appndx:proof-of-thm-two-indp} shows that
%(\ref{eqn:rate_r1_eq1}), (\ref{eqn:rate_r1_eq2}), (\ref{eqn:rate_r2_eq1}) and (\ref{eqn:rate_r2_eq2}) guarantee that
%$\Pr(E_{21,i}''\bigcap E_{20,i}^{'c} \bigcap E_{1,i}^c \bigcap E_{0,i}^c)$ and
%$\Pr(E_{22,i}''\bigcap E_{20,i}^{''c} \bigcap E_{1,i}^c \bigcap E_{0,i}^c)$ can be made arbitrarily small by
%taking $n$ large enough.

This concludes the proof of theorem \ref{thm:achieve}. \tend
%Finally, $\Pr(E_{3,i}\bigcap E_{2,i}^c \bigcap E_{1,i}^c \bigcap E_{0,i}^c)$ can
%be made arbitrarily small as long as (\ref{eqn:rate_r1_eq1}) and (\ref{eqn:rate_r2_eq1}) are satisfied.

%%%%%%%%%%%%%%%%%%%%%%%%%%%%%%%%%%%%%%%%%%%%%%%%%%%%%%%%%%%%%%%%%%%%%%%%%%%%%%%%%%%%%%%%%%%%%%%%%%%%%%%%%%%%%%%%%%%%%%
%%%%%%%%%%%%%%%%%%%%%%%%%%%%%%%%%%%%%%%%%%%%%%%%%%%%%%%%%%%%%%%%%%%%%%%%%%%%%%%%%%%%%%%%%%%%%%%%%%%%%%%%%%%%%%%%%%%%%%

\subsection{An Upper Bound}
\label{sec:general_upper_bound}
\begin{proposition}
\label{thm:upper_bound}
\it
Assume the broadcast channel setup of theorem \ref{thm:achieve}.
Then, for sending independent information, any achievable rate pair $(R_1,R_2)$ must satisfy
\begin{eqnarray*}
    R_1       & \le &    I(X;Y_1) + C_{21},\\
    R_2       & \le &    I(X;Y_2) + C_{12},\\
    R_1 + R_2 & \le &    I(X;Y_1,Y_2),
\end{eqnarray*}
for some distribution $p(x)$ on $\mX$.
\end{proposition}
\smallskip

\begin{proof}
The proof uses the cut-set bound \cite[theorem 14.10.1]{cover-thomas:it-book}. First we define an equivalent
system by introducing two orthogonal channels $X_2' - Y_1'$ from $\Rbad$ to $\Rgood$ and $X_1' - Y_2'$ from
$\Rgood$ to $\Rbad$. The joint probability distribution function then becomes
\[
    p\left((y_1,y_1'), (y_2,y_2') | x,x_1',x_2'\right) = p(y_1,y_2|x)p(y_1'|x_2')p(y_2'|x_1'),
\]
where the signal received at $\Rgood$ is $(Y_1,Y_1')$ and the signal received at $\Rbad$ is $(Y_2,Y_2')$.
As in the proof in section \ref{sec:Pe_analysis_phy_deg}, we select $\mX_1'$, $\mX_2'$,
$\mY_1'$, $\mY_2'$, $p(x_1')$, $p(x_2')$,
$p(y_1'|x_2')$ and $p(y_2'|x_1')$ such that the capacities of the channels
$X_2' - Y_1'$ and $X_1' - Y_2'$ are $C_{21}$ and $C_{12}$ respectively. Additionally, the codewords
for the conference transmissions are determined independently from the source codebook so we set
$p(x,x_1',x_2') = p(x)p(x_1')p(x_2')$.
Now, from the cut-set bound, letting the transmitter and $\Rbad$ form one group and $\Rgood$ the second group,
we have
\begin{eqnarray*}
    R_1 & \le & I(X,X_2';Y_1,Y_1'|X_1') \\
        &  =  & I(X_2';Y_1,Y_1'|X_1') + I(X;Y_1,Y_1'|X_1',X_2')\\
        &  =  & I(X_2';Y_1'|X_1') + I(X_2';Y_1|X_1',Y_1')  \\
        &     & \phantom{xxxxxx} +I(X;Y_1'|X_1',X_2') + I(X;Y_1|X_1',X_2',Y_1')\\
        &  =  & I(X_2';Y_1') + I(X;Y_1)\\
        &  =  &  C_{21} + I(X;Y_1),
\end{eqnarray*}
where $I(X_2';Y_1|X_1',Y_1') = I(X;Y_1'|X_1',X_2') = 0$ follows from direct application of the distribution
function. Similarly we obtain the rate constraint on $R_2$. Lastly, for the sum-rate consider the transmitter
in one group and the receivers in the second. Then, the cut-set bound results in
\begin{eqnarray*}
    R_1 + R_2 & \le & I(X;Y_1,Y_2,Y_1',Y_2'|X_1',X_2')\\
              &  =  & I(X;Y_1,Y_2|X_1',X_2') \\
              &     & \phantom{xxxxxxxx}+ I(X;Y_1',Y_2'|X_1',X_2',Y_1,Y_2)\\
              &  =  & I(X;Y_1,Y_2),
\end{eqnarray*}
yielding the last constraint in the proposition.
\end{proof}

\subsection{Remarks}
\begin{remark}
Observing the rate constraints in theorem \ref{thm:achieve}
%to the rate constraints without cooperation
%in equations (\ref{eqn:marton_R1}) - (\ref{eqn:Marton_SumRate}) we can see that cooperation in general increases
%the available rate.
we can see that when (\ref{eqn:C21_rate conbtraint_Rev}) and (\ref{eqn:C12_rate conbtraint_Rev}) are satisfied
then the cooperative rates are greater than the non-cooperative rates due to the (generally) positive terms adding to $I(U;Y_1)$
and $I(V;Y_2)$.
%
%
%Indeed, comparing the rate constraints for $\Rgood$ we have
%from theorem \ref{thm:achieve_CEG} that
%\begin{eqnarray}
%    R_1 & \le & I(U;Y_1,\hU) \nonumber\\
%\label{eqn:2nd_form_of_R1_constr}
%        & = & I(U;Y_1) + I(U;\hU|Y_1).
%\end{eqnarray}
%The additive term $I(U;\hU|Y_1)$ is non-zero as long as $U - Y_1 - \hU$ do not form a Markov chain, i.e.,
%$\hU$ contains information on $U$ which does not appear already in $Y_1$. This can be done as long as
%$Y_2$ is not a degraded version of $Y_1$.
\end{remark}
\begin{remark}
\label{cmt:card_bound}
We note that although we present a single letter characterization of the rates, we are not able to apply
standard cardinality bounding techniques such as those used in \cite{HajekPursley:79} or \cite{Salehi:78} for
bounding $||\mU||$ and $||\mV||$. The
method of \cite{HajekPursley:79} cannot be applied since it relies on the fact that the auxiliary random variables are
independent, which is not the case here. The method of \cite{Salehi:78} cannot be applied as explained in the
comment for theorem 2 in \cite{HanCosta:87}. The cardinality bounds on $||\mhU||$ and $||\mhV||$ are trivial since they
are transmitted over noiseless links.
\end{remark}
\begin{remark}
The relay strategies can be divided into two general classes. The first class is referred to as {\em
decode-and-forward} (DAF). In this strategy, the relay first decodes the message intended for the destination and
then generates a relay message based on the decoded information. The second class is referred to as
{\em estimate-and-forward} (EAF). In this class the relay does not decode the message intended for the destination
but transmits an estimate of its channel input to the destination.
For the physically degraded BC we used DAF, based on \cite[theorem 1]{CoverG:79},
to derive theorem \ref{thm:converse},
 and for the general BC we used the EAF scheme of \cite[theorem 6]{CoverG:79}, to derive theorem
 \ref{thm:achieve}. Of course, one can also combine
 both strategies and perform partial decoding at each receiver of the other receiver's message before conferencing,
 following \cite[theorem 7]{CoverG:79}. This combination will, in general, result in an increased achievable rate region.
\end{remark}

\subsection{Special Cases}
\subsubsection{No Cooperation: $C_{12} = C_{21} = 0$}
Consider first cooperation from $\Rbad$ to $\Rgood$. Setting $C_{21} = 0$ in theorem \ref{thm:achieve}
%implies that $\hU$ must satisfy $I(\hU;Y_2|U,Y_1) = 0$.
%This condition is equivalent to setting
implies that
\begin{equation}
\label{eqn:special_no_coop}
     H(\hU|Y_1) = H(\hU|Y_2).
\end{equation}
From equation (\ref{eqn:def_RU}), the constraint on $R_1$ can be
written in the form
\begin{eqnarray*}
    R_1 & \le &  I(U;Y_1) + I(U;\hU|Y_1).
\end{eqnarray*}
Now we find $I(U;\hU|Y_1)$:
\begin{eqnarray}
     I(U;\hU|Y_1)   & = & H(\hU|Y_1) - H(\hU|Y_1,U) \nonumber\\
                    & \stackrel{(a)}{=} & H(\hU|Y_2) - H(\hU|Y_1,U) \nonumber\\
\label{eqn:example_no}
                    & \stackrel{(b)}{=} & H(\hU|Y_2,Y_1,U) - H(\hU|Y_1,U) \\
                    & = & -I(\hU;Y_2|Y_1,U).\nonumber
\end{eqnarray}
where (a) is due to (\ref{eqn:special_no_coop}), and (b) is due to the Markov chain
$U - (U,V) - X - (Y_1,Y_2) - Y_2 - \hU$, which implies that given $Y_2$, $\hU$ is independent of
$Y_1$ and $U$.
Now, since mutual information is non-negative, we conclude that $I(U;\hU|Y_1)= 0$.
%Therefore, the minimum in (\ref{eqn:rate_r1_eq1}) is zero and
Hence, the rate constraint on $R_1$ becomes
\[
    R_1 \le I(U;Y_1).
\]
Similarly, the maximum rate $R_2$ is given by $I(V;Y_2)$, and in conclusion when $C_{12} = C_{21} = 0$ we
resort back to the rate region without cooperation derived in \cite{Marton:79} (with a constant $W$).

\subsubsection{Full Cooperation: $C_{12}=H(Y_1|Y_2)$, $C_{21}=H(Y_2|Y_1)$}
When $C_{12} = H(Y_1|Y_2)$, we get from (\ref{eqn:C12_rate conbtraint_Rev}) that
%at the maximal cooperation rate
\begin{eqnarray*}
    H(Y_1|Y_2)   =   C_{12} & \ge & I(\hV;Y_1) - I(\hV;Y_2)\\
                        & = & H(\hV|Y_2) - H(\hV|Y_1),
\end{eqnarray*}
which is satisfied when $\hV = Y_1$.
Plugging this into (\ref{eqn:def_RV}), we get that
when full cooperation from $\Rgood$ to $\Rbad$ is available, the rate constraint for $\Rbad$ becomes
\[
    R_2 \le I(V;Y_2,Y_1).
\]
Using the same reasoning we conclude that when full cooperation from $\Rbad$ to $\Rgood$ is available,
the rate constraint for $\Rgood$ becomes
%\[
    $R_1 \le I(U;Y_1,Y_2)$.
%\]

\subsubsection{Partial Cooperation}
When %using the maximum possible cooperation rate for the
%method of theorem \ref{thm:achieve}, while
$0 < C_{12} < H(Y_1|Y_2)$ and $0 < C_{21} < H(Y_2|Y_1)$,  we get that
\begin{eqnarray}
  C_{21}  & \ge &  H(\hU|Y_1) - H(\hU|Y_2)\nonumber\\
\label{eqn:C21_special}
  \Rightarrow H(\hU|Y_1) & \le & C_{21} + H(\hU|Y_2).
\end{eqnarray}
%We now have the following proposition:
%\begin{proposition}
%    \label{prop:equality}
%    For every $0 \le C_{21} \le H(Y_2|Y_1)$ there exists some probability distribution $p(\hu|y_2)$ such that
%    $C_{21} = H(\hU|Y_1) - H(\hU|Y_2)$.
%\end{proposition}
%The proof for this proposition is given in appendix \ref{appndx:proof_prop}.
%
%Similarly there exists some $p(\hv|y_2)$ such that
%%\[
%  $C_{12} = H(\hV|Y_2) - H(\hV|Y_1)$.
%%\]
%Under these conditions,
Hence,
the achievable rate to $\Rgood$ is upper bounded by
\begin{eqnarray}
    R_1 & \le & I(U;Y_1,\hU) \nonumber \\
        &  =  & I(U;Y_1) + I(U;\hU|Y_1) \nonumber\\
        &  =  & I(U;Y_1) + H(\hU|Y_1) - H(\hU|U,Y_1) \nonumber\\
        &  \stackrel{(a)}{\le}  & I(U;Y_1) + H(\hU|Y_2) - H(\hU|U,Y_1) + C_{21} \nonumber\\
        &  \stackrel{(b)}{=}  & I(U;Y_1) + H(\hU|Y_2,Y_1,U) - H(\hU|U,Y_1) + C_{21} \nonumber\\
\label{eqn:partial_coop_explicit}
    R_1     &  \le  & I(U;Y_1) + C_{21} - I(\hU;Y_2|U,Y_1).
\end{eqnarray}
where (a) is due to (\ref{eqn:C21_special}) and (b) follow from the same reasoning leading to
equation (\ref{eqn:example_no}).
%%relationship $p(\hu|u,v,x,y_1,y_) = p(\hu|y_2)$ defined in theorem \ref{thm:achieve_CEG}.
Similarly, $R_2 \le I(V;Y_2) + C_{12} - I(\hV;Y_1|V,Y_2)$.
%Note that there is not benefit to increasing $C_{21}$ beyond $I(\hU;Y_2) - I(\hU;Y_1)$.

%\begin{remark}
%\label{cmt:upper_general}
%Assume that $C_{ij} < I(Y_i;X|Y_j)$. Note that there exists a gap of $I(\hU;Y_2|U,Y_1)$ and
Note that there exist negative terms  $-I(\hU;Y_2|U,Y_1)$ and   $-I(\hV;Y_1|V,Y_2)$
%between the achievable rates and their respective upper bounds.
in the achievable rate upper bounds.
This can be explained as follows: the mutual information $I(\hU;Y_2|U,Y_1)$ can be considered as a type of
``ancillary'' information that $\hU$ contains, since this information is contained in $\hU$ while $U$ and $Y_1$
are already known - therefore, this information is a ``noise'' part of $Y_2$ which does not include
any helpful information for decoding $U$ at $\Rgood$. Thus, for cooperating in the optimal way, $\hU$ has
to be a type of ``sufficient and complete'' cooperation information.
%\end{remark}
\begin{comment}
        \begin{remark}
            Combining theorem \ref{thm:achieve_CEG} and proposition \ref{prop:equality} we can obtain the following rate expressions:
            \begin{corollary}
                Assume the broadcast channel setup of theorem \ref{thm:achieve_CEG}. Then, there exist mapping functions
                $p(\hu|y_2)$ and $p(\hv|y_1)$ such that the rate pair
                \begin{eqnarray*}
                    R_1 & \le & I(U;Y_1) + C_{21} - I(\hU;Y_2|U,Y_1),\\
                    R_2 & \le & I(V;Y_2) + C_{12} - I(\hV;Y_1|V,Y_2),
                \end{eqnarray*}
                and
                \[
                    C_{12} = I(\hV;Y_1|Y_2), \qquad C_{21} = I(\hU;Y_2|Y_1),
                \]
                where the joint distribution $p(u,v,x,y_1,y_2,\hu,\hv) = p(u,v,x)p(y_1,y_2|x)p(\hu|y_2)p(\hv|y_1)$, and
                $||\mhV|| \le ||\mY_1||$, $||\mhU|| \le ||\mY_2||$, is achievable.
            \end{corollary}
        \end{remark}
        \begin{remark}
            Note that the equality in proposition \ref{prop:equality} should be understood in the asymptotic sense: we
            can approach the value arbitrarily close.
        \end{remark}

\end{comment}

%%%%%%%%%%%%%%%%%%%%%%%%%%%%%%%%%%%%%%%%%%%%%%%%%%%%%%%%%%%%%%%%%%%%%%%%%%%%%%%%%%%%%%%%%%%%%%%%%%%%%%%%%%%%%%%%%%%%%
%%%%%%%%%%%%%%%%%%%%%%%%%%%%%%%%%%%%%%%%%%%%%%%%%%%%%%%%%%%%%%%%%%%%%%%%%%%%%%%%%%%%%%%%%%%%%%%%%%%%%%%%%%%%%%%%%%%%%
%%%%%%%%%%%%%%%%%%%%%%%%%%%%%%%%%%%%%%%%%%%%%%%%%%%%%%%%%%%%%%%%%%%%%%%%%%%%%%%%%%%%%%%%%%%%%%%%%%%%%%%%%%%%%%%%%%%%%

\section{The General Broadcast Channel with a Single Common Message}
\label{sec:commonmsg}

\begin{figure}[ht]
      \center\scalebox{0.48}{\includegraphics{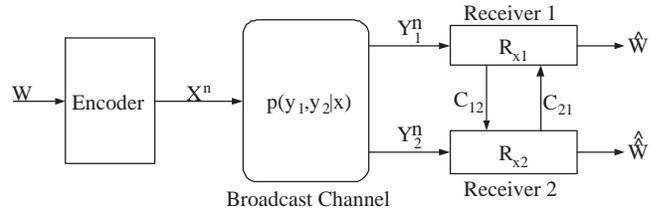}}
%\centerline{\psfig{file=Broadcast_Channel_Common.eps,width=10.5cm,height=3.75cm}}
\caption{\small The single message broadcast channel with cooperating receivers. $\hW$ and
$\hat{\hW}$ are the estimates of $W$ at $\Rgood$ and $\Rbad$ respectively.}
\label{fig:broadcast-cooperation}
\end{figure}

We now consider the case where only a single message, rather than two independent
messages, is transmitted to both receivers. The main motivation for considering this case is that in the two
independent messages
case it is difficult to specify an explicit cooperation scheme, and we therefore have to represent cooperation
through auxiliary random variables. %, for which we are not able to derive cardinality bounds.
Hence, we cannot identify directly the gain from cooperation,
except in the case of full cooperation, and we also cannot evaluate the achievable region.
For the single common message case, we are able to derive results for partial cooperation without auxiliary variables,
{\em which make this region explicitly computable}.
This scenario is depicted in figure \ref{fig:broadcast-cooperation}. %\footnote{Note that in
%figure \ref{fig:broadcast-cooperation} we abuse notation: since each receiver produces its own estimate $\hat{W}$ of $W$,
%both $\hat{W}$'s are not necessarily the same.}.

For this scenario we need to specialize the definitions of a code and the average probability of error as follows:

%\begin{definition}
%\label{def:CommonCode}
\begin{itemize}
\item A $\left(2^{nR},n,(C_{12},C_{21})\right)$ {\em code} for sending a common message
over the broadcast channel with cooperating receivers having conference links
of capacities $C_{12}$ and $C_{21}$ between them,
%consists of a message set $\mW = \left\{1,2,...,2^{nR}\right\}$, an encoding function
%\[
%    f: \mW \mapsto \mX^n,
%\]
%a $\left(C_{12},C_{21}\right)$-admissible conference
%\begin{eqnarray*}
%    & h_{12} : & \mY_1^n \times \mW_{21} \mapsto \mW_{12},\\
%    & h_{21} : & \mY_2^n \times \mW_{12} \mapsto \mW_{21},
%\end{eqnarray*}
%and two decoding functions
%\begin{eqnarray*}
%    & g_1 : & \mW_{21} \times \mY_1^n \mapsto \mW,\\
%    & g_2 : & \mW_{12} \times \mY_2^n \mapsto \mW.
%\end{eqnarray*}
is defined in a similar manner to definition \ref{def:codes} with $\mW_1$, $\mW_2$ and $\mW_1 \times \mW_2$ all replaced
with $\mW = \left\{1, 2, ...,2^{nR}\right\}$.
%\end{definition}

%\begin{definition}
%\label{def:perrCommon}
\item The {\em average probability of error} is defined
similarly to definition \ref{def:perr} with $W_1$ and $W_2$ replaced with $W$.
%as the
%probability that at least one of the receivers does not correctly decode the message:
%\[
%    \Perr = \Pr\left( g_1(W_{21},Y_1^n) \ne W \;\; \mbox{or} \;\; g_2(W_{12},Y_2^n) \ne W\right).
%\]
%We also define the average probability of error for each of the receivers
%as:
%\begin{eqnarray}
%\label{eqn:def_pe1}
%    P_{e1}^{(n)} & = & \mbox{Pr}\left(g_1\left(W_{21},Y_1^n\right) \ne W\right), \\
%\label{eqn:def_pe2}
%    P_{e2}^{(n)} & = & \mbox{Pr}\left(g_2\left(W_{12},Y_2^n\right) \ne W\right).
%\end{eqnarray}
%\end{definition}

\end{itemize}

\begin{figure}[ht]
      \center\scalebox{0.36}{\includegraphics{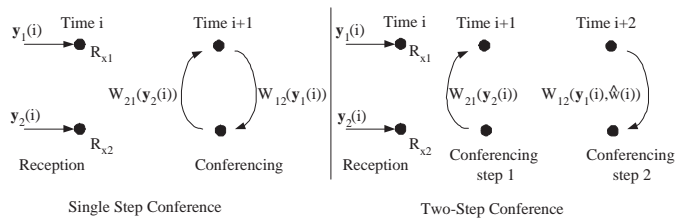}}
%    \centering    {\includegraphics[width = 0.75\textwidth]{OneTwoConference.eps}}
    \caption{\small Schematic description of the single-step and the two-step conference schemes.}
  \label{fig:conference}
\end{figure}

The capacity for the non-cooperative single message scenario is given in \cite{ElGamal:79} by
\begin{equation}
    C = \sup_{p(x)} \Big\{ \min \big( I(X;Y_1) , I(X;Y_2) \big) \Big\}.
\end{equation}
In the following we consider two cooperation schemes, referred to as a single-step scheme and
a two-step scheme. These schemes are described in figure \ref{fig:conference}.
In the single-step scheme, after reception each receiver generates a single cooperation
message based on its channel input. In the two-step scheme, after reception one receiver generates a
cooperation message based only on its channel input, as in the previous case, but the second receiver generates
its cooperation message only after decoding (which is done with the help of the conference message from
the first receiver).
 In both cases each receiver generates a single conference message, however in the single-step conference
 the emphasis is on low delay, while in the two-step conference we sacrifice delay in order to gain rate.

\subsection{Decoding with a Single-Step Cooperation}
\label{subsec:one_delay}
In this section we constrain both decoders to output their decoded messages
after a conference that consists of a single message from each receiver, based only on its received
channel input. For this case, we can specialize the derivation of theorem \ref{thm:achieve}
and get the following achievable rate for the broadcast channel with partially cooperating
receivers:
\begin{theorem}
\label{thm:onedelay}
\it
Let $\left(\mX, p(y_1,y_2|x), \mY_1 \times \mY_2\right)$ be any discrete memoryless broadcast
channel, with cooperating receivers having noiseless conference links of finite capacities
$C_{12}$ and $C_{21}$, as defined in section~\ref{sec:defs}.
%Assume the broadcast channel setup of theorem \ref{lemma:achieve_CEG_special}.
Then, for sending a common message
to both receivers, any rate $R$ satisfying
\begin{eqnarray*}
    R & \le & \sup_{p(x)} \Big\{\min \big\{ I(X;Y_1,\hU), I(X;Y_2,\hV)\big\}\Big\},
\end{eqnarray*}
subject to
\begin{eqnarray*}
    C_{21} & \ge & I(\hat{U};Y_2) - I(\hU;Y_1),\\
    C_{12} & \ge & I(\hat{V};Y_1) - I(\hV;Y_2),
\end{eqnarray*}
for some joint distribution $p(x,y_1,y_2,\hu,\hv) =
p(x)p(y_1,y_2|x)p(\hu|y_2)p(\hv|y_1)$ is achievable, with $||\mhU||
\le ||\mY_2||+1$ and $||\mhV|| \le ||\mY_1||+1$.
\end{theorem}

\smallskip

The proof of theorem \ref{thm:onedelay} follows the same lines of the proof of theorem
\ref{thm:achieve} and will not be repeated here.
%is provided in appendix \ref{appndx:proof-of-lemma-single-step-common}.
%%Note that this result is very intuitive: the usage of $\hU$ in the decoding increases the probability of
%%error due to the uncertainty in $Y_2$ even when $X$ and $Y_1$ are known. This uncertainty is compensated
%%by the cooperation information from the other receiver, contributing to an increase in the rate.
We next show how we can increase the rates by introducing the two-step conference.

\subsection{Decoding with a Two-Step Cooperation}
We consider a two-step conference: at the first step only one receiver decodes the message. The second receiver
decodes after the second step. Therefore, after the first receiver decodes the message, relaying
to the second receiver reduces to the decode-and-forward relay situation of \cite[theorem 1]{CoverG:79}.
The rates achievable with a two step conference
are given in the following theorem:
\begin{theorem}
\label{thm:two_delay}
\it
%Let $\left(\mX, p(y_1,y_2|x), \mY_1 \times \mY_2\right)$ be any discrete memoryless broadcast
%channel, with cooperating receivers having noiseless conference links of finite capacities
%$C_{12}$ and $C_{21}$, as defined in section \ref{sec:defs}.
Assume the broadcast channel setup of theorem \ref{thm:onedelay}.
Then, for sending a common message
to both receivers, any rate $R$ satisfying
\begin{eqnarray*}
    R & \le & \sup_{p(x)}\Bigg[ \max\Big\{ R^{12}(p(x)), R^{21}(p(x)) \Big\} \Bigg]\\
    R^{12}(p(x)) & \triangleq & \min \bigg( I(X;Y_1) + C_{21} , \\
                 &  & \qquad  I(X;Y_2) - I(\hV;Y_1|Y_2,X) \\
                 &  & \qquad +  \min\big(C_{12},H(\hV|Y_2)-H(\hV|Y_1)\big) \bigg),
    \end{eqnarray*}
    \begin{eqnarray*}
    R^{21}(p(x)) & \triangleq & \min \bigg( I(X;Y_2) + C_{12} , \\
                 &  & \qquad  I(X;Y_1) - I(\hU;Y_2|Y_1,X) \\
                 &  & \qquad + \min\big(C_{21},H(\hU|Y_1)-H(\hU|Y_2)\big) \bigg),
\end{eqnarray*}
for some joint distribution $p(x,y_1,y_2,\hu,\hv) =
p(x)p(y_1,y_2|x)p(\hu|y_2)p(\hv|y_1)$ is achievable, with $||\mhU||
\le ||\mY_2||+1$ and $||\mhV|| \le ||\mY_1||+1$, and with the
appropriate $C_{12} \ge I(\hV;Y_1|Y_2,X)$  or $C_{21} \ge
I(\hU;Y_2|Y_1,X)$ (the one used for the first cooperation step).
\end{theorem}
\smallskip
\begin{proof}
\subsubsection{Overview of Coding Strategy}
The scheme described in theorem \ref{thm:onedelay} uses a single-step conference for
both decoders. However, if we let one receiver use a two-step conference,
then that receiver, instead of using conference information derived from the raw input of the other receiver,
can use information generated
by the second receiver after it already decoded the message. This conference information
is less noisy, and thus the rate to the first receiver can be increased.

To put this in more concrete terms,
assume that at time $i+1$, $\Rgood$ sends to $\Rbad$ the index $s'_{i+1}$
of the partition into which its relay message at time $i$, denoted $z_{\hv,i}$, belongs. In appendix
\ref{appndx:proof-of-lemma-single-step-common} we show that
$\Rbad$ can decode the message $w_{0,i}$ with an arbitrarily small probability of error as long as
\begin{eqnarray}
    \label{eqn:rate_bound_for_first_step}
    R & \le & I(X;Y_2) - I(\hV;Y_1|Y_2,X)\nonumber\\
        &  & \qquad + \min\left(C_{12},H(\hV|Y_2)-H(\hV|Y_1)\right),
\end{eqnarray}
and
\begin{equation}
    \label{eqn:capacity _constraint_for_first_step}
    C_{12} \ge I(\hV;Y_1|Y_2,X).
\end{equation}
%Recall that a successful decoding at $\Rbad$ implies that
%$\left(\xvec(w_{0,i}),\yvec_1,\yvec_2(i)\right) \in \typ(X,Y_1,Y_2)$,
%for some $\yvec_1 \in S'_{s'_{i+1}}$. In section \ref{subsec:one_delay} we required that
%$\Rbad$ only decodes $w_{0,i}$. However, in the modified scheme
%we require that $\Rbad$ correctly decodes the pair $(\xvec(w_{0,i}),\yvec_1(i))$.
% We show that this does not impose additional rate constraints. In fact,
% reducing the rate by $H(Y_1|Y_2,X)$ is the loss we suffer in order to
%be able to decode $Y_1^n$ at $\Rbad$. This loss is acceptable only if we can
%compensate for it through the cooperation.
%Therefore, after decoding at $\Rbad$, both $\Rbad$ and $\Rgood$ know $\yvec_1(i)$, and
%thus both know the set $\typ\left(X|\yvec_1(i)\right) \triangleq
%\left\{\xvec \in \mX^n: (\xvec, \yvec_1(i)) \in \typ(X,Y_1)\right\}$ to which $\xvec(w_{0,i})$ belongs.

We now introduce the following modifications to the scheme used
in theorem \ref{thm:onedelay}:
\subsubsection{Relay Sets Generation at $\Rbad$}
    $\Rbad$ partitions the message set $\mW$ into $2^{nC_{21}}$ subsets in a uniform and independent manner.
    Denote these subsets with $\tS''_{\ts''}, \: \ts'' \in \left[1,2^{nC_{21}}\right]$.
\subsubsection{Relay Encoding at $\Rbad$}
    $\Rbad$ has an estimate $\hat{\hw}_{0,i}$ of the message $w_{0,i}$. Now, $\Rbad$ looks for the partition into which
    $\hat{\hw}_{0,i}$ belongs and sends the index of this partition, denoted $\ts''_{i+2}$, to $\Rgood$ at time $i+2$.
\subsubsection{Decoding at $\Rgood$}
    Upon reception of $\yvec_1(i)$, $\Rgood$ generates the set $\mL_1(i) =  \left\{w \in \mW : \left(\xvec(w), \yvec_1(i)\right)
    \in \styp(X,Y_1)\right\}$. At time $i+2$, upon reception of $\ts''_{i+2}$, $\Rgood$ looks for an index $w$ such that
    $w \in \mL_1(i) \bigcap \tS''_{\ts''_{i+2}}$. If a unique such $w$ exists
    then $\Rgood$ sets $\hw_{0,i} = w$, otherwise an error is declared.
\subsubsection{Bounding the Probability of Error}
    Using the proof technique in \cite[theorem 1]{CoverG:79}, it can be easily shown that assuming correct decoding
    at $\Rbad$, then any rate $R \le I(X;Y_1) + C_{21}$ is achievable to $\Rgood$.

Combining the bounds derived above, we conclude that with a two-step conference at $\Rgood$,
any rate satisfying
\begin{eqnarray*}
R      & \le & \min \bigg( I(X;Y_1) + C_{21} ,  I(X;Y_2) - I(\hV;Y_1|Y_2,X) \\
       &   & \qquad \qquad     + \min\big(C_{12},H(\hV|Y_2)-H(\hV|Y_1)\big) \bigg),\\
C_{12} & \ge & I(\hV;Y_1|Y_2,X),
\end{eqnarray*}
ia achievable. Repeating the same derivation when $\Rbad$ uses a two-step conference, and combining with
the previous case proves theorem \ref{thm:two_delay}.
\end{proof}

Setting $\hU = Y_2$, $\hV = Y_1$ in theorem \ref{thm:two_delay} we obtain the following achievable
region:
\begin{corollary}
\label{thm:_cor_two_delay}
{ \it
%\it Let $\left(\mX, p(y_1,y_2|x), \mY_1 \times \mY_2\right)$ be any discrete memoryless broadcast channel, with
%cooperating receivers having noiseless conference links of finite
%capacities $C_{12}$ and $C_{21}$, as defined in section \ref{sec:defs}.
Assume the broadcast channel setup of theorem \ref{thm:onedelay}.
Then, for sending a common message to both receivers, any rate $R$
satisfying
\begin{eqnarray*}
    R & \le & \sup_{p(x)}\Bigg[ \max\Big\{ R^{12}(p(x)), R^{21}(p(x)) \Big\} \Bigg]\\
    R^{12}(p(x))\!\! & \triangleq & \!\!\min \bigg(\! I(X;Y_1) + C_{21} , I(X;Y_2) - H(Y_1|Y_2,X)\\
    &  &   \phantom{xxxxxxx} +  \min\big(C_{12},H(Y_1|Y_2)\big) \bigg),\\
    R^{21}(p(x))\!\! & \triangleq &\!\! \min \bigg(\! I(X;Y_2) + C_{12} , I(X;Y_1)
                    - H(Y_2|Y_1,X) \\
    & & \phantom{xxxxxxx} + \min\big(C_{21},H(Y_2|Y_1)\big) \bigg),
\end{eqnarray*}
with the appropriate $C_{12} > H(Y_1|Y_2,X)$  or  $C_{21} >
H(Y_2|Y_1,X)$ (the one used for the first cooperation step), is
achievable.}
\end{corollary}
This gives a partial cooperation result without auxiliary random variables.

\subsection{An Example for Corollary \ref{thm:_cor_two_delay}}
\begin{figure}[ht]
    \centering
    {\includegraphics[width = 0.42\textwidth]{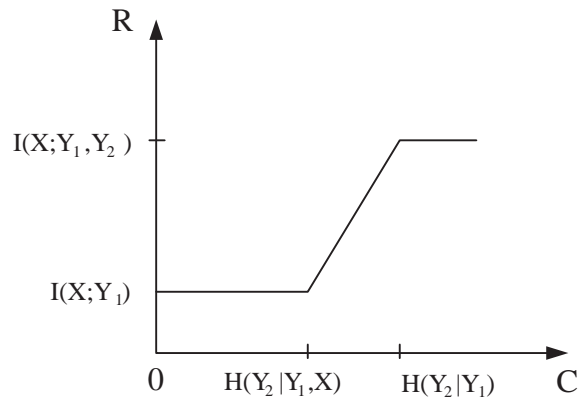}}
    \caption{\small Achievable rate vs. C, for the two independent, identical, BSBCs  with a single common message,
        resulting from corollary \ref{thm:_cor_two_delay}.}
    \label{fig:common_example}
\end{figure}

%\subsubsection{Example for Lemma \ref{lemma:onedelay}}
\label{sec:example_1_common}
Consider two independent, identical,  BSBCs with transition probability $p$, and cooperation
links of capacities $C_{12} = C_{21} = C$. For this case, corollary \ref{thm:_cor_two_delay}
gives the following maximum achievable rate:
\begin{eqnarray*}
    R & = & \sup_{p_0} \bigg\{ \min\Big[H(Y_1) - h(p) + C,\\
        &   & \qquad\qquad \quad  \min\left(H(Y_1) + C , H(Y_1,Y_2) \right) - 2h(p) \Big] \bigg\},\\
    & = & \sup_{p_0} \bigg\{ \min\Big[H(Y_1) - 2h(p) + C, H(Y_1,Y_2)  - 2h(p) \Big] \bigg\},
\end{eqnarray*}
for $C \ge h(p)$, where $\mY_1 = \mY_2 = \mX = \left\{0,1\right\}$, $p_0 = \Pr(X = 0)$, and
\begin{eqnarray*}
\Pr(y_1,y_2) & = &\left\{
    \begin{array}{cl}
        (1-p)^2p_0 + p^2(1-p_0), & y_1 = y_2 = 0\\
        p(1-p),     & y_1 \ne y_2\\
        p^2p_0 + (1-p)^2(1-p_0), & y_1 = y_2 = 1
    \end{array}
    \right.\\
\Pr(y_1) & = & \left\{
    \begin{array}{cl}
        (1-p)p_0 + p(1-p_0), & y_1 = 0\\
        pp_0 + (1-p)(1-p_0), & y_1 = 1.
    \end{array}
    \right.
\end{eqnarray*}
Solving for the supremum for each value of $C$, we get the achievable rates depicted in figure \ref{fig:common_example}.
Note the linear increase in the achievable rate for $H(Y_2|Y_1,X) < C < H(Y_2|Y_1)$.

\subsection{An Upper Bound}
  \label{sec:upper_bound_common}
The upper bound for the single common message case can be obtained from the bound for the two independent messages
 case in proposition \ref{thm:upper_bound}:
  \begin{corollary}
  \label{thm:upper_bound_common}
  \it
  Let $\left(\mX, p(y_1,y_2|x), \mY_1 \times \mY_2\right)$ be any discrete memoryless broadcast
  channel, with cooperating receivers having noiseless conference links of finite capacities
  $C_{12}$ and $C_{21}$, as defined in section~\ref{sec:defs}.
  %Assume the broadcast channel setting of theorem \ref{thm:achieve_CEG_special}.
  Then, for sending a common message to both receivers, any rate $R$ must satisfy
  \begin{eqnarray*}
      R & \le & \sup_{p(x)} \Big\{ \min \big( I(X;Y_1) + C_{21}, I(X;Y_2) + C_{12}, \\
        &   & \qquad \qquad \qquad I(X;Y_1,Y_2) \big) \Big\}.
  \end{eqnarray*}
  \end{corollary}
  \begin{proof}
  Follows directly from proposition \ref{thm:upper_bound} by noting that the common rate has to satisfy all three
  constraints: the individual rates and the sum rate.
  \end{proof}
\subsection{Remarks}
\begin{remark}
Note that there are special cases where the lower bound of corollary \ref{thm:_cor_two_delay} coincides with
the upper bound of corollary \ref{thm:upper_bound_common}, yielding the capacity for these cases.
For example, assume a strong version of the ``more capable" condition of \cite{ElGamal:79}:
 $I(X;Y_1) >> I(X;Y_2)$\footnote{The precise condition
 requires that $I(X;Y_1)  > I(X;Y_2) + C_{12} - C_{21} + H(Y_2|Y_1,X)$ for all input distributions $p(x)$.}
  for all input distributions $p(x)$ on $\mX$ .
 Assume also that  $H(Y_2|Y_1,X) < C_{21} < H(Y_2|Y_1)$ and
$H(Y_1|Y_2,X) < C_{12} < H(Y_1|Y_2)$.
Under these conditions, we have that $I(X;Y_1) + C_{21} > I(X;Y_2) + C_{12} - H(Y_1|Y_2,X)$.
Thus, if $\Rgood$ is helping $\Rbad$ first, the achievable rate is $I(X;Y_2) + C_{12} - H(Y_1|Y_2,X)$.
If $\Rbad$ is helping $\Rgood$ first, then the achievable rate is $I(X;Y_2) + C_{12}$.
Since $C_{12} - H(Y_1|Y_2,X) < C_{12}$, this cooperation scheme achieves the upper bound
$R = \sup_{p(x)} \left\{I(X;Y_2) + C_{12}\right\}$.
\end{remark}
\begin{remark}
Note that the capacity region for the deterministic broadcast channel with cooperating receivers follows from
corollary \ref{thm:_cor_two_delay} and corollary \ref{thm:upper_bound_common}. This region was derived in \cite{Draper:ISIT04}.
For this case we have that
$H(Y_1|X) = H(Y_2|X) = 0$ hence $I(X;Y_i) = H(Y_i)$, $i=1,2$.
%Setting $\hU = Y_2$ and $\hV = Y_1$ in lemma \ref{lemma:onedelay}
The achievable rate (from corollary \ref{thm:_cor_two_delay}) is given by
\begin{eqnarray*}
    R & \le & \min\left\{ H(Y_2) + C_{12}, H(Y_1) + \min\left( C_{21}, H(Y_2|Y_1) \right) \right\}\\
        &  =  & \min \left\{H(Y_2) + C_{12} , H(Y_1) + C_{21}, H(Y_1,Y_2)\right\},
\end{eqnarray*}
and the same from corollary \ref{thm:upper_bound_common}.
\end{remark}

\begin{remark}
    We note that although the expressions in \eqref{eqn:rate_bound_for_first_step} and
    \eqref{eqn:capacity _constraint_for_first_step} seem different from the EAF expression of \cite[theorem 6]{CoverG:79},
    given in theorem \ref{thm:onedelay}
    (cf. $R \le I(X;Y_2,\hV), \mbox{ subject to } C_{12}  \ge  I(\hat{V};Y_1) - I(\hV;Y_2)$), this does not improve on
    the achievable rate of the standard EAF. The reason is that every rate achievable according to
    \eqref{eqn:rate_bound_for_first_step}-\eqref{eqn:capacity _constraint_for_first_step} can also be
    achieved with the standard EAF using the same mapping of the auxiliary RV and an appropriate time-sharing\footnote{This
    observation is due to Shlomo Shamai and Gerhard Kramer.}.
    However, when considering {\em a specific, fixed assignment of the auxiliary random variable} (such as in corollary
    \ref{thm:_cor_two_delay}) then the rate achievable with \eqref{eqn:rate_bound_for_first_step}-\eqref{eqn:capacity _constraint_for_first_step}
    is indeed greater than the classic EAF with the same assignment.

\end{remark}

\section{Conclusions}
\label{sec:conclude}
In this paper we investigated the effect of cooperation  between receivers on the rates for the broadcast channel.
As communication networks evolve, it can be expected that in future networks, nodes that are close enough to be able
to communicate directly, will use this ability to help each other in reception.
Accommodating this characteristic, we extended the traditional
broadcast scenario, in which each decoder is assumed to operate independently, into a scenario where
the receivers have finite capacity links used for cooperation.
We analyzed three related scenarios: the physically degraded BC - for which we derived the capacity region,
the general BC for which we presented an achievability result, and the single common message
case. For the last case we identified a special case where capacity can be achieved.
We note that it is not trivial to extend these results to more than two steps, since the intermediate
steps need to extract information from partial relay information. Although this can be done by introducing additional auxiliary
variables, obtaining a computable region is not a simple task.
This study is an initial step in this investigation and future work includes several extensions:
a natural first extension is to consider a fully wireless system, and extend the analysis to the Gaussian case.
 Another extension is to consider the interaction between the Wyner-Ziv compression and the achievable rates for the general channel.

\section*{Acknowledgements}
The authors would like to thank the reviewers for their careful reading of the manuscript and their useful suggestions.

\useRomanappendicesfalse
\appendices
\setcounter{equation}{0}
\renewcommand{\theequation}{\thesection.\arabic{equation}}

\section{Background Results}
\label{Appndx_BoundSet}
Consider the construction in section \ref{sec:achieve}.
Let $\mathcal{L}(i-1) = \left\{w_{2}: \left({\bf y}_2(i-1), {\bf u}(w_{2}|s_{i-1})\right)
 \in A_{\epsilon}^{(n)}\right\}$.
We bound $E_{\yvec_2}\left\{||\mathcal{L}(i-1)||  \right\}$. Let,
   \begin{displaymath}
   \psi\left(w_2 \left|\yvec_2(i-1) \right.\right) = \left\{
   \begin{array}{cc}
   1, & \left(\uvec(w_2|s_{i-1}), \yvec_2(i-1) \right) \in \typ \\
   0, & \mbox{otherwise.}
   \end{array}
   \right.
   \end{displaymath}
    Hence, as in \cite[theorem 1]{CoverG:79}, we can write the random variable $||\mL(i-1)||$ as a sum of random
    variables:
    \[
        ||\mathcal{L}(i-1)|| = \sum_{w_2 = 1}^{2^{nR_2}} \psi\left(w_2 \left|\yvec_2(i-1) \right.\right),
    \]
    and therefore
    \begin{eqnarray*}
        E_{\yvec_2}\Big\{||\mathcal{L}(i-1)||   \Big\} & = &  E_{\yvec_2}\left\{\psi\left(w_{2,i-1} \left|\yvec_2(i-1) \right.\right)
        \right\}\\
        &  & \phantom{x} +\!\!\! \sum_{\substack{w_2=1 \\w_2\ne w_{2,i-1}}}^{2^{nR_2}} E_{\yvec_2}\left\{\psi\left(w_2 \left|\yvec_2(i-1) \right.\right)\right\}.
    \end{eqnarray*}
   When $w_2 \ne w_{2,i-1}$ we get from the properties of independent sequence
    (\cite[theorem 8.6.1]{cover-thomas:it-book})   that
    %as long as $s_{i-1}$ is correct, then
    \begin{eqnarray*}
        E_{\yvec_2}\left\{\psi\left(w_2 \left|\yvec_2(i-1) \right.\right)\right\}
         & =  & \Pr\left\{\psi\left(w_2 \left|\yvec_2(i-1) \right.\right)=1\right\}\\
         & \le &  2^{-n\left(I(U;Y_2)-3\epsilon \right)},
    \end{eqnarray*}
    thus,
    \begin{equation}
    \label{eqn:Dec_set_bound}
        E_{\yvec_2}\Big\{||\mathcal{L}(i-1)||   \Big\}  \le   1+ 2^{nR_2} 2^{-n\left(I(U;Y_2)-3\epsilon \right)}.
    \end{equation}
  Note that this result holds also when considering the strongly typical set rather than the weakly typical
  set.

\setcounter{equation}{0}
\section{Proof of the Achievable Rate to the First Decoder in Theorem \ref{thm:two_delay} (equations \eqref{eqn:rate_bound_for_first_step}
and \eqref{eqn:capacity _constraint_for_first_step})}
\label{appndx:proof-of-lemma-single-step-common}

\subsection{Overview of Coding Strategy}
The encoder generates a single codebook in a random and independent manner. Next, the first relay partitions
its collection of relay codewords ($\mZ(\hV)$ for $\Rgood$) % or $\mZ(\hU)$ for $\Rbad$)
into disjoint sets. When a channel
input is received, the first relay finds the index of the partition set which contains a relay codeword
jointly typical with its channel input, and
transmits it over the noiseless conference link to the second receiver. Then, the second receiver looks for
a unique source codeword that is jointly typical with its channel input, and with at least one of the relay codewords
in the set of possible codewords received from the first relay.
%Note that here we use {\em joint} decoding and therefore obtain superior rates.

In the following analysis we assume that $\Rgood$ is the first relay and $\Rbad$ decodes first.

\subsection{Codebook Generation and Encoding at the Transmitter}
Fix $p(x)$ and
generate $2^{nR}$ i.i.d. codewords $\xvec$, with
$p(\xvec(w)) = \prod_{i=1}^n p(x_i(w))$, $w \in \mW =  \left\{1,2,..., 2^{nR}\right\}$.
%Denote this codebook by $\mC$.
%
%\subsubsection{Encoding at the Transmitter}
For transmitting the message $w_{0,i}$ at time $i$, the transmitter outputs $\xvec(w_{0,i})$ to the channel.

\subsection{Relay Sets Generation}
Fix $p(\hv|y_1)$.
\begin{itemize}
\item Consider the p.d.f. $p(\hv) = \sum_{\mX, \mY_1, \mY_2} p(\hv|y_1)p(y_1,y_2|x)p(x)$ on $\mhV$.
\item $\Rgood$ generates $2^{n R_1'}$ $\hvvec$ sequences in an i.i.d. manner according
to $p(\hvvec(z_{\hv})) = \prod_{i=1}^n p(\hv_i(z_{\hv}))$, $z_{\hv} \in \mZ(\hV) = \left\{1,2,...,2^{nR_1'}\right\}$.

\item $\Rgood$ partitions the message set $\mZ(\hV)$ into $2^{nC_{12}}$ sets, by assigning an
index between $\left[1, 2^{nC_{12}}\right]$ to each $z_{\hv} \in \mZ(\hV)$, in a random, independent
and uniform manner over $\left[1, 2^{nC_{12}}\right]$. Denote these sets by $S'_{s'}$,
$s' \in \left[1, 2^{nC_{12}}\right]$.
%\item $\Rbad$ partitions the set of typical $Y_2^n$ sequences, $\typ(Y_2)$, into $2^{nC_{21}}$ sets, by assigning an
%index between $\left[1, 2^{nC_{21}}\right]$ to each typical $\yvec_2 \in \typ(Y_2)$, in a random, independent
%and uniform manner over $\left[1, 2^{nC_{21}}\right]$. Denote these sets by $S''_{s''}$,
%$s'' \in \left[1, 2^{nC_{21}}\right]$.
\end{itemize}
%The generation of the set of $\huvec$ relay codewords for $\Rbad$ (indexed by $z_{\hu} \in \mZ(\hU)$) and its partition (denoted $S''_{s''}$) is done in
%a parallel manner to that described for $\Rgood$, w.r.t. the probability $p(\hu)$ on $\mhU$.
\subsection{Decoding and Encoding at the Relay ($\Rgood$)}
\begin{itemize}
\item Upon reception of $\yvec_1(i)$, the relay $\Rgood$ decides that  $z_{\hv,i}\in \mZ(\hV)$ was received if
$\left(\hvvec(z_{\hv,i}),\yvec_1(i)\right) \in \styp(\hV,Y_1)$. Now, $\Rgood$
finds the index $s_{i+1}'$ of the set
$S_{s_{i+1}'}'$ s.t. $z_{\hv,i} \in S_{s_{i+1}'}'$. Then, at time
$i+1$, $\Rgood$ transmits $s_{i+1}'$ to $\Rbad$ through the finite
capacity noiseless conference link. If there is no $z_{\hv} \in \mZ(\hV)$ such that $\hvvec(z_{\hv})$ is jointly typical
with $\yvec_1(i)$, an error is declared.
%\item Upon reception of $\yvec_2(i-1)$, the relay $\Rbad$ finds the index $s_i''$ of the set
%$S_{s_i''}''$ s.t. $\yvec_2(i-1) \in S_{s_i''}''$. Then, at time $i$, $\Rbad$ transmits $s_i''$ to
%$\Rgood$ over the finite capacity noiseless conference link. If $\yvec_2(i-1) \notin \typ(Y_2)$, an error
%is declared.
\end{itemize}
%The relay decoding and encoding at $\Rbad$ is done in a parallel manner to that at $\Rgood$.

\subsection{Decoding the Source Message at $\Rbad$}
At the $i$'th transmission interval $\Rbad$ generates the set
$\mL_2(i) = \left\{w \in \mW:
\left(\xvec(w),\yvec_2(i)\right) \in \styp(X,Y_2)\right\}$. At the
$(i+1)$'th transmission interval, $\Rbad$ receives $s'_{i+1}$
from $\Rgood$ through the noiseless conference link. $\Rbad$ then
looks for a unique $\hw_0$ s.t. $\hw_0 \in \mL_2(i)$ and
$\exists z_{\hv} \in S'_{s'_{i+1}}$, for which
$\left(\xvec(\hw_0),\yvec_2(i), \hvvec(z_{\hv}) \right) \in
\styp(X,Y_2,\hV)$. If such  unique $\hw_0$ exists, then
$\hw_0$ is the decoded message at time $i$. If there is none, or
there is more than one, an error is declared.
%%\subsubsection{Decoding at $\Rbad$}
%%At the $i$'th transmission interval, $\Rbad$ receives $s'_i$ from $\Rgood$ through the noiseless conference link.
%%$\Rbad$ then looks for a unique $\xvec(\hw_0)$ s.t. $\exists \yvec_1 \in S'_{s'_i}$, for which
%%$\left(\xvec(\hw_0), \yvec_1, \yvec_2(i-1)  \right) \in \typ(X,Y_1,Y_2)$.
%%If such a unique $\xvec(\hw_0)$ exists, then $\hw_0$ is the decoded message. If there is none, or there is more than
%%one, an error is declared.
%Decoding at $\Rbad$ is done in a parallel manner to the decoding at
%$\Rgood$.

\subsection{Analysis of the Probability of Error}
%We now analyze the average probability of error for this scheme.

\subsubsection{Error Events}
The error events for the scheme described above, for decoding the message $w_{0,i}$, are:
\begin{enumerate}
\item Relay decoding fails: \\%$E_{0,i} = E_{0,i}' \bigcup E_{0,i}''$\\
%    $E'_{0,i} = \left\{\nexists z_{\hu} \in \mZ(\hU) \mbox{ s.t. } (\huvec(z_{\hu}), \yvec_2(i)) \in \styp(\hU,Y_2)  \right\}$,\\
    $E_{0,i} = \Big\{\nexists z_{\hv} \in \mZ(\hV) \mbox{ s.t. }$ \phantom{xxxxxxxxxxx}\\
       \phantom{xxxxxxxxxxxxxx} $ (\hvvec(z_{\hv}), \yvec_1(i)) \in \styp(\hV,Y_1)  \Big\}$.
\item Joint typicality decoding fails: Let $E_{1,i} = E'_{1,i} \bigcup E''_{1,i}$, where\\
    $E'_{1,i} = \left\{\left(\xvec(w_{0,i}), \yvec_1(i), \yvec_2(i) \right) \notin \styp(X,Y_1,Y_2)\right\}$,\\
%    $E''_{1,i} = \left\{\left(\xvec(w_{0,i}), \yvec_1(i), \huvec(z_{\hu,i}) \right) \notin \styp(X,Y_1,\hU)\right\}$,\\
    $E''_{1,i} = \left\{\left(\xvec(w_{0,i}), \hvvec(z_{\hv,i}), \yvec_2(i) \right) \notin \styp(X,\hV,Y_2)\right\}$.
%\item Decoding at $\Rgood$ fails:\\
%    $E_{2,i} = E_{2,i}' \bigcup E_{2,i}''$,\\
%    $E_{2,i}' = \left\{ \nexists z_{\hu} \in S''_{s''_{i+1}} \; \mbox{for which} \; \left(\xvec(w_{0,i}), \yvec_1(i), \huvec(z_{\hu})\right) \in \right. $
%        $ \left.   \styp(X,Y_1,\hU)   \right\},$\\
%    $E_{2,i}'' = \left\{\exists w \ne w_{0,i}, w\in \mL_1(i) \; \mbox{s.t.} \; \exists z_{\hu} \in S''_{s''_{i+1}}, \right. $
%        $\left. \left(\xvec(w),\yvec_1(i), \huvec(z_{\hu}) \right) \in \styp(X,Y_1,\hU)\right\}$.
\item Decoding at $\Rbad$ fails:
    $E_{2,i} = E_{2,i}' \bigcup E_{2,i}''$,\\
    $E_{2,i}' = \Big\{ \nexists z_{\hv} \in S'_{s'_{i+1}} \; \mbox{for which}$\phantom{xxxxxxxxxxxxxx}\\
        \phantom{xxxxxxxx}        $ \; \left(\xvec(w_{0,i}), \hvvec(z_{\hv}), \yvec_2(i)\right) \in  \styp(X,\hV,Y_2)   \Big\},$\\
    $E_{2,i}'' = \Big\{\exists w \ne w_{0,i}, w \in \mL_2(i) \; \mbox{s.t.} \; \exists z_{\hv} \in S'_{s'_{i+1}}, $\phantom{xxxxx}\\
      \phantom{xxxxxxxxxx}  $ \left(\xvec(w),\hvvec(z_{\hv}), \yvec_2(i) \right) \in\styp(X,\hV,Y_2)\Big\}$.
\end{enumerate}
%where $w_{0,i}$ denotes the message transmitted at time $i$, and
%$\mL_2(i) = \left\{w: \left(\xvec(w),\yvec_2(i)\right) \in \styp(X,Y_2)\right\}$.
Next, applying the union bound we get that
\begin{eqnarray*}
    \Perr & = & \Pr\left(\bigcup_{k=0}^2 E_{k,i} \right) \\
    & =  &  \Pr(E_{0,i}) + \Pr\left(E_{1,i}\bigcap E_{0,i}^c\right)\\
    &     & \phantom{xxxxxxxxx}   +  \Pr\left(E_{2,i}\bigcap E_{1,i}^c\bigcap E_{0,i}^c\right).
\end{eqnarray*}
%where the inequality is due to the union bound.

\subsubsection{Bounding the Probabilities of the Error Events}
\label{subsec:bounds_pe_common}
 Following the same argument as in section \ref{sec:dec_rly_2},
%$R_2' \ge I(\hU;Y_2)$ implies that we can make $\Pr(E'_{0,i}) \le \frac{\eps}{2}$ for
%arbitrarily small $\eps > 0$, by taking $n$ large enough. Similarly,
$R_1' \ge I(\hV;Y_1)$ implies that
 taking $n$ large enough,
we can make %$\Pr(E_{0,i}) \le \frac{\eps}{2}$ and hence
$\Pr(E_{0,i}) \le \eps$.
Next, from the properties of strongly typical sequences (see \cite[lemma 13.6.1]{cover-thomas:it-book}),
 by taking $n$ large enough, we can make $\Pr(E'_{1,i}) \le \frac{\eps}{2}$. Additionally, the
 Markov lemma, \cite[lemma 4.2]{BetgerLecNotes} implies that we can make
 $\Pr(E''_{1,i} \bigcap E_{1,i}'^{c}\bigcap E_{0,i}^{c}) \le \frac{\eps}{2}$
% and $\Pr(E'''_{1,i} \bigcap E_{0,i}^{''c}) \le \frac{\eps}{3}$
for any arbitrary $\eps >0$ by taking $n$ large enough.
Therefore, by the union bound, $\Pr(E_{1,i}\bigcap E_{0,i}^{c}) \le \eps$.
We also have that $\Pr(E'_{2,i} \bigcap E_{1,i}^c \bigcap E_{0,i}^c)  = 0$
because under $E_{1,i}^c\bigcap E_{0,i}^c$ we have that $\xvec(w_{0,i}),\yvec_2(i)$ and $\hvvec(z_{\hv,i})$ are jointly typical, and
by construction, $z_{\hv,i} \in S'_{s'_{i+1}}$.
%Following the same reasoning, we also have that $\Pr(E'_{3,i} \bigcap E_{1,i}^c \bigcap E_{0,i}^c) = 0$.
Hence, we need to show that the probability $\Pr(E''_{2,i}\bigcap E_{1,i}^c \bigcap E_{0,i}^c)$
%and $\Pr(E''_{3,i} \bigcap E_{1,i}^c \bigcap E_{0,i}^c)$
can be made
arbitrarily small. Note that due to the symmetry of the construction, the probability of error does not depend
on the specific message $w_{0,i}$ transmitted.

\subsubsection{Bounding $\Pr(E''_{2,i} \bigcap E_{1,i}^c \bigcap E_{0,i}^c)$}
The probability of $E''_{2,i} \bigcap E_{1,i}^c \bigcap E_{0,i}^c$ can be written as
{
\setlength{\arraycolsep}{0mm}
\begin{eqnarray*}
& & \Pr(E''_{2,i}\bigcap E_{1,i}^c \bigcap E_{0,i}^c) \\
 & &\; =  \Pr\Big(\Big\{\exists z_{\hv} \in S'_{s'_{i+1}}, \exists w \ne w_{0,i}, w\in\mL_2(i),\\
 & &  \phantom{llll}          \left(\xvec(w), \yvec_2(i),\hvvec(z_{\hv})\right)\in \styp(X,Y_2,\hV)\Big\}\bigcap E_{1,i}^c \bigcap E_{0,i}^c\Big)\\
%%%%%%%%%%%%%%%%%%%%
&  & \;\stackrel{(a)}{=}     \Pr\Big(\Big\{\exists w \ne w_{0,i}, w\in\mL_2(i),\\
&  & \phantom{lll} \left(\xvec(w), \yvec_2(i),\hvvec(z_{\hv,i})\right)\in \styp(X,Y_2,\hV)\Big\}\bigcap E_{1,i}^c\bigcap E_{0,i}^c\Big)\\
&   & \phantom{xxx}  + \Pr\Big( \Big\{\exists w \ne w_{0,i}, w\in\mL_2(i), \exists z_{\hv} \in S'_{s'_{i+1}},
        z_{\hv} \ne z_{\hv,i},\\
&  & \phantom{xlll}  \left(\xvec(w), \yvec_2(i),\hvvec(z_{\hv})\right)\in \styp(X,Y_2,\hV)\Big\}\bigcap E_{1,i}^c\bigcap E_{0,i}^c\Big)\\
&   & \;\triangleq \Pr(E''_{2,1,i}) + \Pr(E''_{2,2,i}),
\end{eqnarray*}
}
where (a) is because the elements of $S'_{s'_{i+1}}$ are selected in an independent manner.

We first bound $\Pr\left(E_{2,1,i}'' \right)$
%Repeating essentially the same steps leading to (\ref{eqn:bound_pE3ai}) we obtain
 as follows:
{\setlength\arraycolsep{0.0cm}
\begin{eqnarray*}
&  & \Pr(E''_{2,1,i})  =   \\
&  & \; \sum_{\mL_2(i)}\Pr\Big( \Big\{\exists w \ne w_{0,i}, w \in \mL_2(i), \left(\xvec(w),\yvec_2(i),\hvvec(z_{\hv,i})\right)\\
&  & \qquad \quad    \in   \styp(X,Y_2,\hV)\Big\} \bigcap E_{1,i}^c \bigcap E_{0,i}^c\Big| \mL_2(i)\Big) \Pr\left(\mL_2(i)\right)\\
%%%%%%%%%%%%%%%%%%%%%%%%%%%%%%%%%%%%%%
&  & \stackrel{(a)}{\le} \;
%    \sum_{\yvec_1(i)}\sum_{\substack{w \in \mL_1(i) \\ w \ne w_{0,i}}}
%    \Pr\Big( \Big\{ \left(\xvec(w),\yvec_1(i),\yvec_2(i)\right)\in \\
%     &  & \qquad \qquad \qquad \typ(X,Y_1,Y_2) \Big\} \bigcap E_0^c \Big| \yvec_1(i)\Big)  \Pr\left(\yvec_1(i)\right)\\
%%%%%%%%%%%%%%%%%%%%%%%%%%%%%%%%%%%%
    E_{\yvec_2}\Bigg\{ \sum_{\substack{w \in \mL_2(i) \\ w \ne w_{0,i}}}\Pr\Big( \Big\{\left(\xvec(w),\yvec_2(i),\hvvec(z_{\hv,i})\right)\in\\
    &  &\qquad\qquad\qquad\qquad  \styp(X,Y_2,\hV)\Big\} \bigcap E_{1,i}^c \bigcap E_{0,i}^c\Big| \yvec_2(i)\Big)\Bigg\}\\
%%%%%%%%%%%%%%%%%%%%%%%%%%%%%%%%%%%%
&  & = E_{\yvec_2}\left\{ \sum_{\substack{w \in \mL_2(i) \\ w \ne w_{0,i}}} \sum_{\substack{\hvvec \in \\
        \styp(\hV|\yvec_2(i),\xvec(w))}}\Pr\left(\hvvec|\yvec_2(i),\xvec(w)\right)\right\}\\
%%%%%%%%%%%%%%%%%%%%%%%%%%%%%%%%%%%%
&  & \stackrel{(b)}{=} E_{\yvec_2}\left\{ \sum_{\substack{w \in \mL_2(i) \\ w \ne w_{0,i}}}
        \sum_{\hvvec \in \styp(\hV|\yvec_2(i),\xvec(w))}
        \Pr\left(\hvvec|\yvec_2(i)\right)\right\}\\
%%%%%%%%%%%%%%%%%%%%%%%%%%%%%%%%%%%%
&  & \le E_{\yvec_2}\Bigg\{ \sum_{\substack{w \in \mL_2(i) \\ w \ne w_{0,i}}} ||\styp(\hV|\yvec_2(i),\xvec(w))||\times\\
    &  &\qquad  \qquad\qquad  \qquad\max_{\substack{\hvvec: \\ \left(\yvec_2(i),\hvvec\right)\in\styp(Y_2,\hV)}}
        \Big\{\Pr\left(\hvvec|\yvec_2(i)\right)\Big\}\Bigg\}\\
%%%%%%%%%%%%%%%%%%%%%%%%%%%%%%%%%%%%
&  & \stackrel{(c)}{\le} E_{\yvec_2}\left\{ \sum_{\substack{w \in \mL_2(i) \\ w \ne w_{0,i}}}
        2^{n(H(\hV | Y_2, X) + 2\eta)}2^{-n\left(H(\hV|Y_2)-2\eta\right)}\right\}\\
%%%%%%%%%%%%%%%%%%%%%%%%%%%%%%%%%%%%%
&  & \le  E_{\yvec_2}\left\{||\mL_2(i)||\right\} 2^{-n\left(H(\hV|Y_2) - H(\hV | Y_2, X) - 4\eta\right)},
\end{eqnarray*}
}
where (a) is because $\mL_2(i)$ is a deterministic function of
$\yvec_2(i)$ and we also applied the union bound and (b) is because $\hvvec(z_{\hv,i})$ is
independent of $\xvec(w)$ for $w \ne w_{0,i}$. The bounds in (c) on the size of the conditionally typical set and
the maximum conditional probability follow from \cite[theorem 5.2]{YeungBook}
with $\eta \rightarrow 0$ as $\eps \rightarrow 0$, assuming that $n$ is large enough.
Lastly we note that here
\[
   \Pr(\yvec_2(i)) \triangleq \Pr\left(\yvec_2(i)\;\mbox{received}\; \big|\; \xvec(w_{0,i}) \;\mbox{transmitted}\right).
\]
Next, applying the same technique to bound the expectation of $||\mL_2(i)||$ as in \cite[theorem 1]{CoverG:79} (see also
derivation of equation (\ref{eqn:Dec_set_bound})), we
get that for $n$ large enough,
\begin{equation}
\label{eqn:Dec_set_bound_common}
    E_{\yvec_2}\left\{||\mL_2(i)||\right\} \le 1 + 2^{n\left(R - I(X;Y_2) + 3\eta\right)}.
\end{equation}
Plugging this back into the bound on $\Pr\left(E''_{2,1,i}\right)$ we get that
{
\begin{eqnarray}
\!\!\!\!\!\!\!\!\!\!\!\!\Pr\left(E''_{2,1,i}\right) & \le & 2^{-n\left(I\left(X;\hV|Y_2\right)-4\eta\right)}\nonumber\\
&  &  + 2^{n\left(R - I(X;Y_2) - H(\hV|Y_2) + H(\hV | Y_2, X) + 7\eta\right)},
\end{eqnarray}
}
which can be made less than any arbitrary $\eps > 0$ by taking $n$ large enough,
as long as\footnote{We assume that $I(X;\hV|Y_2)>0$ otherwise the relay message does not help decoding the source message
at $\Rbad$.}
\begin{equation}
\label{eqn:common_bound_1}
    R \le I(X;Y_2) - H(\hV | Y_2, X) + H(\hV|Y_2).
\end{equation}
For bounding $\Pr(E''_{2,2,i})$ we
%follow the same steps leading to (\ref{eqn:bound_E2bi}) and obtain
begin essentially in the same manner and get that
{%\setlength\arraycolsep{0.0cm}
\begin{eqnarray*}
&  & \!\!\!\!\!\!\!\!   \Pr(E''_{2,2,i}) \\
    &  \stackrel{(a)}{\le} &  E_{\yvec_2,\hvvec}
      \Bigg\{ \Pr\left(\Big\{\exists w \ne w_{0,i}, w\in \mL_2(i), \exists z_{\hv} \in S'_{s'_{i+1}},\right.\\
    &  & \qquad \qquad \qquad
         z_{\hv} \ne z_{\hv,i},\left(\xvec(w),\yvec_2(i),\hvvec(z_{\hv})\right) \in \\
    &  & \qquad\qquad \qquad  \qquad \styp(X,Y_2,\hV)\Big\} \Big|\yvec_2(i),\hvvec(z_{\hv,i})\Big)  \Bigg\}\\
%&  & \; \stackrel{(b)}{\le} E_{\yvec_1,\yvec_2}\Bigg\{ \sum_{\substack{\yvec_2 \in S''_{s''_{i+1}} \\ \yvec_2 \ne \yvec_2(i)}}
%    \Pr\Big(\exists w \ne w_0, w\in \mL_1(i), \\
%&  & \qquad \quad \left.\left.\left(\xvec(w),\yvec_1(i),\yvec_2\right) \in \typ(X,Y_1,Y_2)  \right|\yvec_1(i),\yvec_2(i)\right)\Bigg\}\\
%%%%%%%%%%%%%%%%%%%%%%%%%
&   \stackrel{(b)}{\le} & E_{\yvec_2,\hvvec}
    \Bigg\{ \sum_{\substack{z_{\hv} \in S'_{s'_{i+1}} \\ z_{\hv} \ne z_{\hv,i}}}
    \sum_{\substack{w \in \mL_2(i) \\ w \ne w_{0,i}}}
    \Pr\Big(\left(\xvec(w),\yvec_2(i),\hvvec(z_{\hv})\right)\in  \\
&   &  \qquad\qquad \qquad \qquad  \quad  \styp(X,Y_2,\hV) \Big| \yvec_2(i),\hvvec(z_{\hv,i})\Big)\Bigg\}\\
%%%%%%%%%%%%%%%%%%%%%%%%%
%&  & \; \le E_{\yvec_1,\yvec_2}\Bigg\{ \sum_{\substack{\yvec_2 \in S''_{s''_{i+1}} \\ \yvec_2 \ne \yvec_2(i)}}
%    \sum_{\substack{w \in \mL_1(i) \\ w \ne w_0}}   \sum_{\substack{\yvec_2 \in \\ \typ(Y_2| \\ \yvec_1(i),\xvec(w))}}
%    \Pr\left(\yvec_2|\yvec_1(i),\xvec(w)\right)\Bigg\}\\
%%%%%%%%%%%%%%%%%%%%%%%%%
&  \stackrel{(c)}{=} &
    E_{\yvec_2,\hvvec}\Bigg\{ \sum_{\substack{z_{\hv} \in S'_{s'_{i+1}} \\ z_{\hv} \ne z_{\hv,i}}}
      \sum_{\substack{w \in \mL_2(i) \\ w \ne w_{0,i}}} \sum_{\hvvec \in \styp(\hV|\yvec_2(i),\xvec(w))}
\Pr\left(\hvvec\right)\Bigg\}\\
%%%%%%%%%%%%%%%%%%%%%%%%%
%&  & \; \le E_{\yvec_1,\yvec_2}\left\{ \sum_{\substack{\yvec_2 \in S''_{s''_{i+1}} \\ \yvec_2 \ne \yvec_2(i)}}
%    \sum_{\substack{w \in \mL_1(i) \\ w \ne w_0}} 2^{n\left(H(Y_2|Y_1,X)+2\eps\right)}2^{-n\left(H(Y_2)-\eps\right)}\right\}\\
%%%%%%%%%%%%%%%%%%%%%%%%%
&   \stackrel{(d)}{\le} & E_{\hvvec}\left\{||S'_{s'_{i+1}} ||\right\} E_{\yvec_2}\left\{||\mL_2(i)||\right\}2^{-n\left(H(\hV) - H(\hV|Y_2,X)-3\eta\right)}\\
%%%%%%%%%%%%%%%%%%%%%%%%%
&   \stackrel{(e)}{\le}& \left( 1 + 2^{n\left(R_1' - C_{12} \right)}\right)\left(1 + 2^{n\left(R - I(X;Y_2) + 3\eta\right)}\right)\times\\
&  &     \qquad\qquad\qquad \qquad \qquad \quad  \quad     2^{-n\left(H(\hV) - H(\hV|Y_2,X)-3\eta\right)}\\
%%%%%%%%%%%%%%%%%%%%%%%%%
%&  & \; = 2^{-n\left(C_{21} - H(Y_2|Y_1,X)-4\eps\right)}\left(1 + 2^{n\left(R - I(X;Y_1) + 3\eps\right)}\right)\\
%%%%%%%%%%%%%%%%%%%%%%%%%
&   \le  &2^{-n\left(C_{12} + H(\hV) - R_1' - H(\hV|Y_2,X)-3\eta\right)}  \\
&  & \phantom{xx}+ 2^{n\left(R - I(X;Y_2) - I(\hV;Y_2,X)+ 6\eta\right)} + 2^{-n\left(I(\hV;Y_2,X)-3\eta\right)}\\
&  & \; \phantom{XX}  + 2^{n\left(R - I(X;Y_2) - C_{12} + R_1' - H(\hV) + H(\hV|Y_2,X)  + 6\eta\right)},
\end{eqnarray*}
}
where (a) is because we dropped the intersection with $E_{1,i}^c \bigcap E_{0,i}^c$,
(b) is due to the union bound,
(c) is because $\hvvec(z_{\hv})$ is independent of $\xvec(w)$ and $\yvec_2(i)$ when $z_{\hv} \ne z_{\hv,i}$, and (d)
is because
\begin{eqnarray*}
&  &     E_{\yvec_2,\hvvec}\big\{||\mL_2(i)||\cdot||S'_{s'_{i+1}}||  \big\}\\
    &  & \phantom{xxxxxxxxxxxx} = E_{\yvec_2}\Big\{E_{\hvvec|\yvec_2}\big\{||\mL_2(i)||\cdot||S'_{s'_{i+1}}||\big\}\Big\}\\
    &  & \phantom{xxxxxxxxxxxx} \stackrel{(f)}{=} E_{\yvec_2}\Big\{||\mL_2(i)||E_{\hvvec|\yvec_2}\big\{||S'_{s'_{i+1}}||   \big\}\Big\}\\
    &  & \phantom{xxxxxxxxxxxx} \stackrel{(g)}{=}  E_{\yvec_2}\Big\{||\mL_2(i)||E_{\hvvec}\big\{||S'_{s'_{i+1}}||   \big\}\Big\}\\
    &  & \phantom{xxxxxxxxxxxx} = E_{\yvec_2}\Big\{||\mL_2(i)||\Big\}E_{\hvvec}\Big\{||S'_{s'_{i+1}}||\Big\},
\end{eqnarray*}
where (f) is because the average size of $\mL_2(i)$ does not depend on $\hvvec(z_{\hv,i})$ when $\yvec_2(i)$ is given, and
(g) is because the average size of $S'_{s'_{i+1}}$ does not depend of $\yvec_2(i)$.
The bounds
on $\Pr(\hvvec)$ and $||\styp(\hV|\yvec_2,\xvec)||$ in (d) follow from \cite[Ch. 5]{YeungBook}.
The bound on $E_{\yvec_2}\left\{||\mL_2(i)||\right\}$ in (e) follows from equation (\ref{eqn:Dec_set_bound_common}).
We note that here
\begin{eqnarray*}
   &  & \!\!\!\!\!\Pr(\yvec_2(i),\hvvec(z_{\hv,i})) \triangleq \Pr\Big(\left(\yvec_2(i), \hvvec(z_{\hv,i})\right) \;\mbox{received}\;
    \big|\; \xvec(w_{0,i})\\
    &  &  \phantom{xxxxxxxxxxxxxxxxxxxxxxxxxxxxxxxxx}\;\mbox{transmitted}\Big).
\end{eqnarray*}
\begin{comment}
    We now prove the bound on $E_{\huvec}\left\{||S''_{s''_{i+1}} ||\right\}$ in (e). Define %first
    \[
        \varphi_m(z_{\hu}) = \left\{
        \begin{array}{cc}
            1, & z_{\hu} \in S_m''\\
            0, & z_{\hu} \notin S_m''.
        \end{array}
        \right.
    \]
    Then, $||S_i''||  = \sum_{z_{\hu}=1}^{2^{nR_2'}} \varphi_i(\huvec)$, and therefore,
    \begin{eqnarray*}
         E\left\{||S''_{s''_{i+1}} ||\right\}
        &  =  & \sum_{z_{\hu}=1}^{2^{nR_2'}}
                 E\left\{\varphi_{s''_{i+1}}(z_{\hu})\right\}\\
        &  =  &    1 +  \sum_{\substack{z_{\hu} =1, z_{\hu} \ne z_{\hu,i}}}^{2^{nR_2'}} \Pr\left(z_{\hu} \in S_{s''_{i+1}}''\right)\\
    %    & \le &    1 + ||\mZ(\hU)||2^{-nC_{21}}\\
        &  \le  &    1 + 2^{n(R_2' - C_{21})},
    \end{eqnarray*}
    %where in the last equality we used $||\mZ(\hU)|| = 2^{nR'}$.
\end{comment}
We conclude that $\Pr\left(E_{2,2,i}''\right)$ can be made smaller than any $\eps > 0$ by taking $n$ large enough,
as long as
\begin{eqnarray}
\label{eqn:common_bound_2}
\!\!\!\!\!\!\!\!\!    R      & \le & I(X;Y_2) - H(\hV|Y_2,X) + C_{12}-R_1'+H(\hV) \\
\label{eqn:common_bound_2_on_C}
\!\!\!\!\!\!\!\!\!    R_1' & \le & C_{12} - H(\hV|Y_2,X) + H(\hV)\\
\label{eqn:common_bound_2.5_on_C}
\!\!\!\!\!\!\!\!\!    R    &  \le & I(X;Y_2) + I(\hV;Y_2,X)\\
\label{eqn:common_bound_3_on_C}
\!\!\!\!\!\!\!\!\!    R_1' & \ge & I(\hV;Y_1),
\end{eqnarray}
where (\ref{eqn:common_bound_3_on_C}) follows from appendix \ref{subsec:bounds_pe_common}.

Now note that making $\Pr(E''_{2,i}\bigcap E_{1,i}^c \bigcap E_{0,i}^c)$ arbitrarily small requires making both
$\Pr(E''_{2,1,i})$ and  $\Pr(E''_{2,2,i})$ arbitrarily small. Thus we also need to satisfy (\ref{eqn:common_bound_1}).
Combining with (\ref{eqn:common_bound_2.5_on_C}) we see that (\ref{eqn:common_bound_1}) guarantees (\ref{eqn:common_bound_2.5_on_C})
and we  are left with (\ref{eqn:common_bound_1}), (\ref{eqn:common_bound_2}), (\ref{eqn:common_bound_2_on_C}) and (\ref{eqn:common_bound_3_on_C}).

The maximum rate is achieved for the minimal $R_1'$, therefore we plug $R_1' = I(\hV;Y_1)$ in
(\ref{eqn:common_bound_2}) and combining with (\ref{eqn:common_bound_1}) we obtain the following achievable rate
\begin{eqnarray}
\label{eqn:common_bound_4_on_C}
    R &  \le & I(X;Y_2) - H(\hV|Y_2,X)  \nonumber\\
        &  & \qquad \quad  + \min\left(C_{12} + H(\hV|Y_1),H(\hV|Y_2)\right).
\end{eqnarray}
From the combination of (\ref{eqn:common_bound_2_on_C}) and (\ref{eqn:common_bound_3_on_C}), we conclude
that this is achievable as long as
\begin{eqnarray}
\label{eqn:common_bound_5_on_C}
    C_{12}  &\ge  & I(\hV;Y_1) + H(\hV|Y_2,X) - H(\hV)\nonumber\\
     &   = &  H(\hV|Y_2,X) - H(\hV|Y_1)\nonumber\\
     &   = & I(\hV;Y_1|X,Y_2).
\end{eqnarray}
%Using similar considerations we can bound $\Pr\left(E''_{2,1,i}\right)$ by\footnote{
%Full details of the derivation will
%appear in the journal version of this paper.}
%\begin{eqnarray}
%\Pr\left(E''_{2,1,i}\right) & \le & 2^{-n\left(I\left(X;Y_2|Y_1\right)-4\eps\right)} \nonumber\\
%&  & + 2^{n\left(R - I(X;Y_1) - H(Y_2|Y_1) + H(Y_2 | Y_1, X) + 7\eps\right)},
%\end{eqnarray}
%which can be made less than any arbitrary $\eps > 0$ by taking $n$ large enough,
%as long as\footnote{We assume that $I(X;Y_2|Y_1)>0$ otherwise the channel is degraded.}
%\begin{equation}
%\label{eqn:common_bound_1}
%    R \le I(X;Y_1) + H(Y_2|Y_1) - H(Y_2 | Y_1, X).
%\end{equation}
 Equations (\ref{eqn:common_bound_4_on_C}) and (\ref{eqn:common_bound_5_on_C}) give the conditions for the
message $W$ to be decoded at $\Rbad$ with an arbitrarily small probability of error by
taking $n$ large enough.
%, as long as
%\begin{equation}
%\label{eqn:common_rate_combined_bound}
%    R \le I(X;Y_1) - H(\hU|Y_1,X) + \min\left(C_{21} , H(\hU | Y_1)\right).
%\end{equation}
%%But then, if this rate is less than the rate without cooperation (that is, if in order to resolve
%%the uncertainty caused by not knowing the received $\yvec_2(i)$, we need to transmit more cooperation information
%%than what can be supported by the finite capacity conference link, i.e. $C_{21} < H(Y_2|Y_1,X)$),
%%we better not use cooperation at all. In this case the maximum allowed rate to $\Rgood$ is bounded by $I(X;Y_1)$.
%%Combining this observation with (\ref{eqn:common_rate_combined_bound}) we get the first term in the minimum
%% in lemma \ref{lemma:onedelay}.
Note that the requirement in (\ref{eqn:common_bound_5_on_C}) implies that when
$C_{12} < I(\hV;Y_1|Y_2,X)$, $\Rgood$ cannot use this cooperation scheme, and the rate to $\Rbad$ is simply
 $I(X;Y_2)$. Combining this with equation (\ref{eqn:common_bound_4_on_C}) yields the rate expression
 in \eqref{eqn:rate_bound_for_first_step} and \eqref{eqn:capacity _constraint_for_first_step}.
% for $R_1(p(x))$ in theorem~\ref{thm:onedelay}.

%Repeating these considerations for $\Rbad$, we get the expression for $R_2(p(x))$.
%The maximum achievable rate is then the supremum over all possible distributions
% $p(x)$.

%%\bibliographystyle{plain}
%%\bibliography{library}

\begin{biography}{Ron Dabora}
received his B.Sc. and M.Sc. degrees in electrical engineering in
1994 and 2000 respectively, from Tel-Aviv University, Tel-Aviv,
Israel. From 1994 to 2000, he was with the Signal Corps of Israel
Defense Forces, and from 2000 to 2003, he was a member of the
Algorithms Development Group, Millimetrix Broadband Networks,
Israel. Since 2003 he is a Ph.D. student at Cornell University,
Ithaca, NY.
\end{biography}

\begin{biography}{Sergio D. Servetto}
was born in Argentina, on January 18, 1968.  He received a
Licenciatura en Inform\'atica from Universidad Nacional de La Plata
(UNLP, Argentina) in 1992, and the M.Sc. degree in Electrical
Engineering and the Ph.D. degree in Computer Science from the
University of Illinois at Urbana-Champaign (UIUC), in 1996 and 1999.
Between 1999 and 2001, he worked at the \'Ecole Polytechnique
F\'ed\'erale de Lausanne (EPFL), Lausanne, Switzerland.  Since Fall
2001, he has been an Assistant Professor in the School of Electrical
and Computer Engineering at Cornell University, and a member of the
fields of Applied Mathematics and Computer Science.  He was the
recipient of the 1998 Ray Ozzie Fellowship, given to ``outstanding
graduate students in Computer Science,'' and of the 1999 David J.\
Kuck Outstanding Thesis Award, for the best doctoral dissertation of
the year, both from the Dept.\ of Computer Science at UIUC.  He was
also the recipient of a 2003 NSF CAREER Award.  His research
interests are centered around information theoretic aspects of
networked systems, with a current emphasis on problems that arise in
the context of large-scale sensor networks.
\end{biography}
\end{document}